\documentclass[useAMS,usenatbib]{mn2e}

\usepackage{graphicx,epsfig}  

\def\etal{\rm et~al.\rm}

\title[Paper II: Calibration and characteristics of the
  Swift UVOT]{Paper II: Calibration of
  the {\it Swift} ultraviolet/optical telescope}

\author[A. A. Breeveld, and 
  UVOT team members]{A. A. Breeveld$^{1}$\thanks{E-mail: aab@mssl.ucl.ac.uk},
P. A. Curran$^{1}$,
E. A. Hoversten$^{2}$,
S. Koch$^{2}$,
W. Landsman$^{3}$,  \and 
F. E. Marshall$^{3}$, 
M. J. Page$^{1}$,
T. S. Poole$^{1}$, 
P. Roming$^{2}$, 
P. J. Smith$^{1}$,  \and
M. Still$^{1,13}$, 
V. Yershov$^{1}$, 
A. J. Blustin$^{1,9}$,
P. J. Brown$^{2,8}$,  
C. Gronwall$^{2}$,  \and
S. T. Holland$^{3,4,5}$,
N. P. M. Kuin$^{1}$,    
K. McGowan$^{1,14}$,  
S. Rosen$^{1,7}$,
P. Boyd$^{3}$,   \and 
P. Broos$^{2}$,
M. Carter$^{1}$, 
M. M. Chester$^{2}$,    
B. Hancock$^{1}$, 
H. Huckle$^{1}$,   \and 
S. Immler$^{3}$,
M. Ivanushkina$^{2,12}$, 
T. Kennedy$^{1}$, 
K. O. Mason$^{1,6}$, 
A. N. Morgan$^{2,10}$,   \and 
S. Oates$^{1}$,
M. de Pasquale$^{1}$,
P. Schady$^{1}$,
M. Siegel$^{2}$,
and D. Vanden Berk$^{2,11}$,  \\
$^{1}$Mullard Space Science Laboratory, University College London, Holmbury St. Mary, Dorking, Surrey, RH5 6NT, UK\\
$^{2}$Department of Astronomy \& Astrophysics, Penn State University, 525 Davey Laboratory, University Park, PA 16802, USA\\
$^{3}$NASA/Goddard Space Flight Center, Greenbelt, MD 20771, USA \\
$^{4}$Universities Space Research Association, 10211 Wincopin Circle, Suite 500, Columbia, MD 21044, USA\\
$^{5}$Centre for Research and Exploration in Space Science and Technology, NASA/Goddard Space Flight Center, Greenbelt, MD 20771, USA \\
$^{6}$Science \& technology Facilities Council, Polaris House, North Star Avenue, Swindon, Wilts SN2 1SZ, UK \\
$^{7}$Department of Phyiscs and Astronomy, University of Leicester, University Road, Leicester, LE1 7RH, UK \\
$^{8}$Department of Phyiscs and Astronomy, University of Utah, Salt Lake City, UT 84112, USA \\
$^{9}$Institute of Astronomy, University of Cambridge, Madingley Road, Cambridge, CB3 0HA, UK \\
$^{10}$Astronomy Department, University of California,  Berkeley, CA 94720-3411, USA \\
$^{11}$Physics Department, St. Vincent College,  Latrobe, PA 15650, USA \\
$^{12}$Northrop Grumman Corp., Aerospace Systems, Vehicle Systems Integration, One Hornet Way, El Segundo, CA 90245, USA \\
$^{13}$NASA Ames Research Center, Moffett Field, CA 94035, USA \\
$^{14}$School of Physics and Astronomy, University of Southampton, Southampton, SO17 1BJ, UK }

\begin{document}

\date{Accepted ??? Received ????; in original form 2009 December 16}

\pagerange{\pageref{firstpage}--\pageref{lastpage}} \pubyear{}

\maketitle

\label{firstpage}

\begin{abstract}
The Ultraviolet/Optical Telescope (UVOT) is one of three instruments onboard the {\it Swift} observatory. The photometric calibration has been published, and this paper follows up with  details on other aspects of the calibration including a measurement of the point spread function with an assessment of the orbital variation and the effect on photometry. A correction for large scale variations in sensitivity over the field of view is described, as well as a model of the coincidence loss which is used to assess the coincidence correction in extended regions. We have provided a correction for the detector distortion and measured the resulting internal astrometric accuracy of the UVOT, also giving the absolute accuracy with respect to the International Celestial Reference System. We have compiled statistics on the background count rates, and discuss the sources of the background, including instrumental scattered light.  In each case we describe any impact on UVOT measurements, whether any correction is applied in the standard pipeline data processing or whether further steps are recommended. 
\end{abstract}


\section{Introduction}
The Ultraviolet/Optical Telescope \citep[UVOT;][]{Roming05} is one of three instruments which make up the {\it Swift} observatory \citep{GN2004} along with the 15--150~keV Burst Alert Telescope \citep[BAT;][]{BS2005} and the 0.2--10~keV X-Ray Telescope \citep[XRT;][]{BDN2005}. {\em Swift} is primarily a gamma-ray burst (GRB) observatory, but is increasingly being used to look at a wide range of targets including active galactic nuclei, supernovae and X-ray transients. 

The UVOT obtains ultraviolet and optical data in parallel with the BAT and the XRT. It incorporates a modified Ritchey--Chr\'{e}tien telescope with a $17 \times 17$ arcmin field-of-view and covers the wavelength range 1600--8000\AA. A filter wheel carries seven broadband filters: three in the optical range ({\em v, b {\em and} u}), three in the UV ({\em uvw1, uvm2 {\em and} uvw2}) and one clear filter ({\it white}) covering the whole wavelength range. In addition there are two grisms, a magnifier and a blocked filter in the filter wheel. 

The UVOT uses a fast readout, micro-channel-plate (MCP) intensified, photon-counting CCD detector with $256 \times 256$ active pixels. Each pixel is subdivided by use of an onboard centroiding algorithm into $8 \times 8$ subpixels, giving a full field of $2048 \times 2048$ subpixels, each of which subtends 0.502 arcsec on the sky after correcting for distortion. Throughout this paper the word `pixel' refers to one of these 0.5 arcsec subpixels, unless it is specifically called a CCD pixel. The full field is read out in 11.0329 ms and each photon may be time-tagged with this timing resolution; this is known as event mode. Alternatively, and more usually, image mode is used, where the photons are built up into an image on board to reduce telemetry. The UVOT is suitable for viewing sources from $\sim~10.5$ to $\sim~23.5$ mag, depending on the filter used and the nature of the source. Further details on the UVOT detector may be found in Section 2 of Poole et al. (2008, hereafter Paper I).

The Swift Data Center\footnote{http://swift.gsfc.nasa.gov/docs/swift/sdc/} at Goddard Space Flight Center  is responsible for processing raw telemetry data from the {\it Swift} satellite and making them available to the HEASARC\footnote{http://heasarc.gsfc.nasa.gov/} for distribution to the public as reduced data products. The standard processing is known as `the pipeline'; the current release is 3.13.17 (August 3rd, 2009).  The Swift Science Center\footnote{http://swift.gsfc.nasa.gov/docs/swift/swiftsc.html} provides a suite of software tools to the HEASARC suitable for analysing or reprocessing the data and applying the calibration (UVOT specific \texttt{FTOOLS} are released as part of HEAsoft {\it Swift} software package\footnote{HEAsoft software can be found at \newline
http://heasarc.gsfc.nasa.gov/docs/software/lheasoft/}), as well as maintaining the Calibration database (CalDB\footnote{CalDB files and associated documentation can be found at \newline 
http://swift.gsfc.nasa.gov/docs/heasarc/caldb/swift/}; the current release of the UVOT CalDB is version 20090930).  

The preliminary in-orbit calibration of UVOT was described in \citet{spie1} and \citet{spie2}. With this paper and Paper I we present a more detailed analysis supported by more data and superceding the earlier work. In Paper I we presented the photometry calibration. Here in Paper II we  include the following: measurements of the point spread function and the effect of the orbit (Section~\ref{SecPSF}); a model of how coincidence loss is affected by high backgrounds (Section~\ref{SecCoi}); mod-8 noise and how it can be removed from images (Section~\ref{SecMod8}); small and large scale positional uniformity (Sections~\ref{SecSSS} and \ref{SecLSS}); the accuracy of UVOT astrometry both relative and absolute (Section~\ref{SecAstrometry})  and  background count rates (Section~\ref{SecBackground}), including scattered light. In each case we assess any impact on UVOT science, describe any correction that is applied in the standard pipeline data processing (and how good this correction is), or whether further steps are recommended under certain circumstances. 

The grism calibration will be published in a separate paper.

\section{Point Spread Function}
\label{SecPSF}

Each broad band filter is associated with its own characteristic point spread function (PSF) with the UV filters tending to broader PSFs; this is due to higher energy photons producing higher energy electrons in the detector photocathode, which then travel further laterally through the detector. The UVOT PSF narrows with high count rates due to the effect of coincidence; at very high count rates the PSF is highly distorted. Thus, the stars used for the measurement of the PSF are of moderate or low count rates. The geometric distortion, as described in Section~\ref{SecDistortion}, would affect the PSF especially if measured away from the centre of an image. All the PSF measurements have therefore been performed on images corrected for distortion. There are several other factors that affect the shape of an individual PSF, (see Sections~\ref{SecPSFVar} and \ref{SecMod8}), thus the PSFs described here are representative examples for each filter. 

Since the aperture used to determine the zero-points in Paper I was 5 arcsec in radius, we have retained the normalization as equal to 1.0 at 5 arcsec to agree with the photometric calibration; 5 arcsec contains approximately 85 per cent of the total flux. 

\subsection{Curve of Growth}
The PSF measurement was done in two stages: the core (up to 5 arcsec) and the wings (from 5 to 30 arcsec). For the cores it was relatively easy to find single exposures of moderate count rate stars with sufficient signal to noise, thus avoiding any need to sum up exposures, which could have blurred the PSFs.  However, to obtain sufficient signal to noise in the wings it was necessary to sum many exposures.  When measuring the PSFs, we used unbinned images to maintain full spatial resolution. 

\begin{figure}
\begin{centering}
\psfig{figure=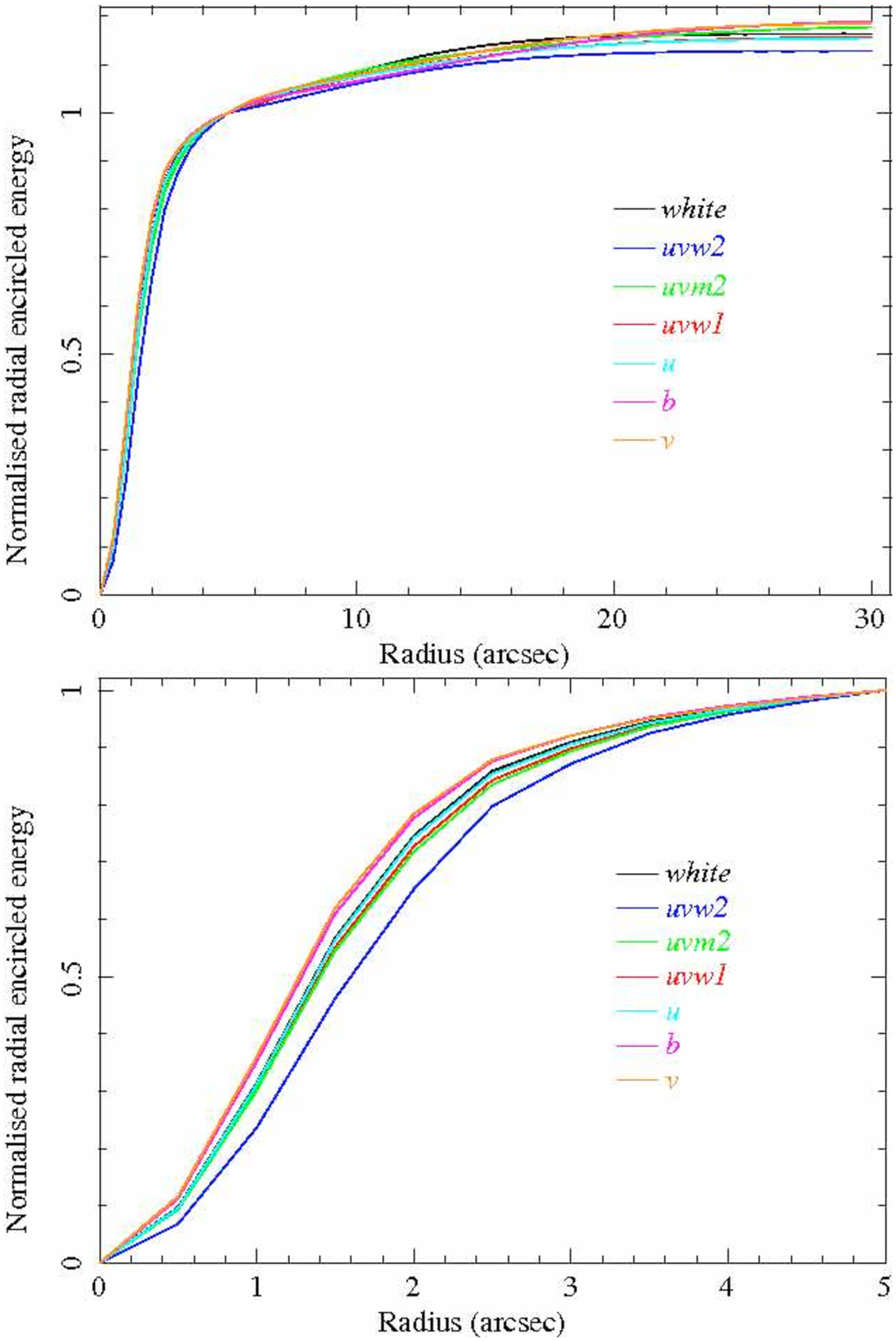,width=60mm,angle=0}
\caption{The curves of growth for the UVOT broadband filters normalised to 1.0 at 5 arcsec. The lower panel shows the cores in more detail. }
\label{FigNormCOG}
\end{centering}
\end{figure}

For the cores a number of single exposures with long exposure times were acquired from the UVOT archive for each filter. Within these fields we identified point-like sources with count rates ranging from 10--20~s$^{-1}$. From these we chose 6--20 relatively isolated sources with which to calculate the PSF. For the wings of the PSFs, we used summed observations of GRB fields for {\it b, v} and {\it white}, while observations of the Chandra Deep Field South (CDF-S, \citealt{Hoversten})  provided deep exposures in each of the {\it u, uvw1, uvm2} and {\it uvw2} filters.  The depth of the combined exposures allows the detail in the wings to be significant above the background out to a radius of 30 arcsec.  For each filter we identified one isolated source with a count rate of between 0.1 and 5~s$^{-1}$. Nearby neighbours to the PSF objects would lead to difficulties in estimating the background level.

For both the core and the wings the DAOPHOT package \citep{Stetson1987} within IRAF was used for the PSF fitting. An analytical Moffat model plus look up table PSF was created from the selected sources using a 15 or 30 arcsec radius for the cores and wings, respectively.  The temporary PSF was then subtracted from nearby sources to improve the field, and the PSF was recalculated. This final analytical PSF was then subtracted from other stars in the field to test the goodness via the residuals. 

We assume that the PSF is radially symmetric (the ellipticity is less than 10 per cent) and calculate the Full Width at Half Maximum (FWHM) from the average sigma parameter of the Moffat fit to the cores, converted to arcsec (see Table~\ref{TabPSF}).  The PSFs of both the cores and the wings were integrated over their radii to convert to curves of growth (COGs) and these COGs were normalised to 1.0 at 5 arcsec. The measured wing COGs were fitted with a sum of two Gaussians centred at zero radius. The renormalised, fitted wing COGs were combined with those of the cores, to give the final COG from 0 to 30 arcsec (see Fig.~\ref{FigNormCOG}). 

\begin{table}
\caption{The FWHM of the Point Spread Function for the UVOT filters. }
\label{TabPSF}
\begin{center}
\begin{tabular}{@{}lccc}
\hline
Filter & FWHM \\
 & (arcsec) \\
\hline
{\it v} & 2.18 \\
{\it b} & 2.19 \\
{\it u} & 2.37 \\
{\it uvw1} & 2.37 \\
{\it uvm2} & 2.45 \\
{\it uvw2} & 2.92 \\
{\it white} & 2.31 \\
\hline
\end{tabular}
\end{center}
\end{table}

\subsection{PSF variations and the effect on photometry}
\label{SecPSFVar}

\begin{figure}
\begin{centering}
\includegraphics[width=84mm]{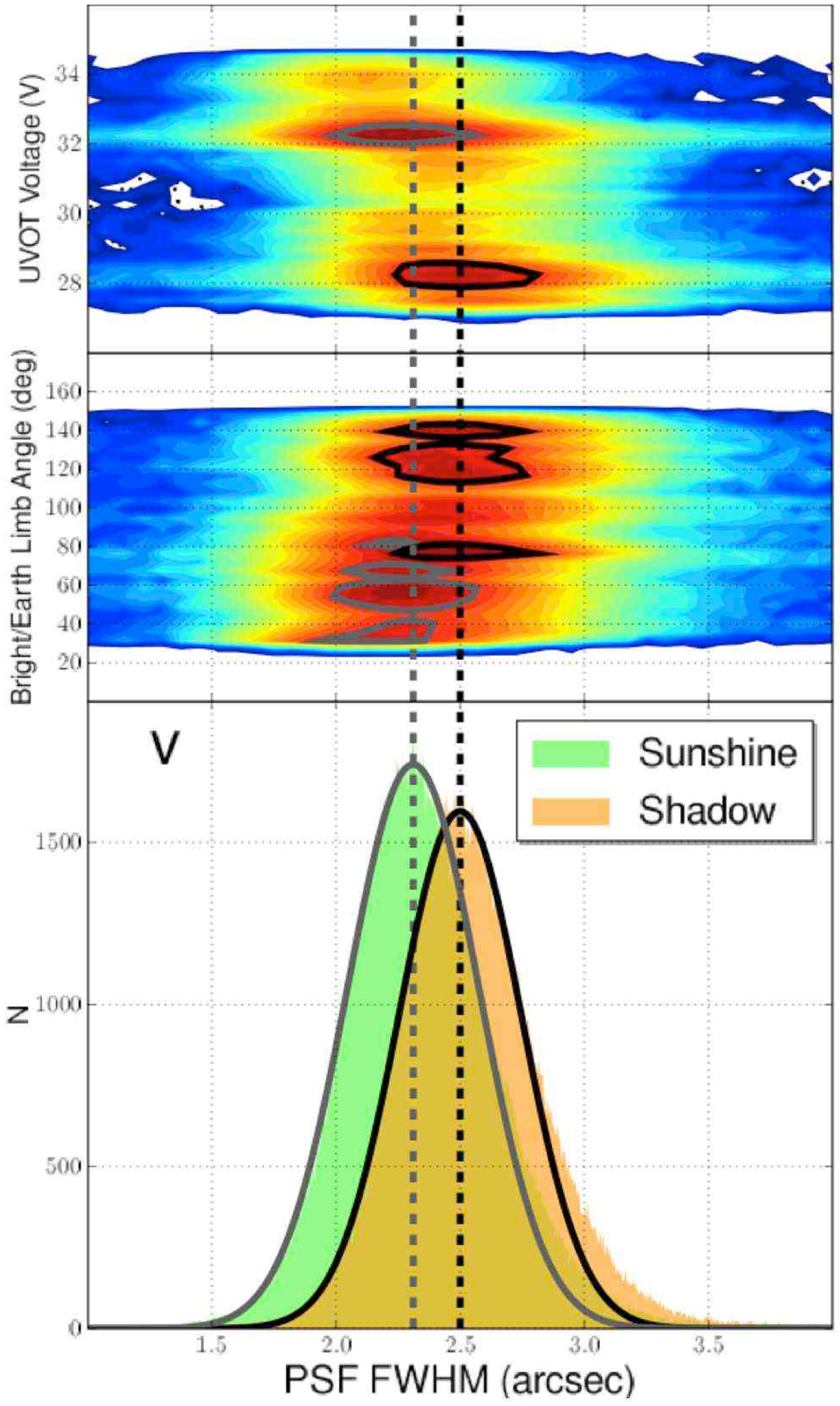}
\caption{PSF FWHM measurements in the {\it v} filter of 579\,766 point sources, using a simple Gaussian fit. The lower panel reveals that the widths of the sources divide into two populations, coordinated according to the spacecraft's location within direct sunlight or Earth shadow. This weak bimodality is uncorrelated with the angle subtended by the spacecraft boresight and the Earth-limb (middle panel), indicating that the effect is unlikely to be caused by external heating. However the UVOT focus heater power depends on the spacecraft voltage, and the PSF width is correlated with the focus heater power (upper panel; in sunlight the voltage is high, in eclipse the spacecraft runs on battery power and the voltage is reduced).}
\label{FigSunShadow}
\end{centering}
\end{figure}

By measuring the FWHM  of many thousands of sources in all filters throughout the mission we are able to show that there has been no significant change in the PSF with time. Nevertheless, we described in Paper I how the PSF changes slightly during the {\it Swift} orbit because the UVOT telescope temperature changes as the satellite moves into and out of the Earth's shadow.  We have used the trending data to show that the width of the PSF does not vary smoothly through the orbit but exhibits two different states depending on whether the spacecraft is in sunshine or in shadow behind the Earth (see Fig.~\ref{FigSunShadow}),  leading to a range in average orbital FWHM in the {\it v} filter of between 2.3 and 2.5 arcsec (8 per cent). In these measurements the FWHM is higher than that given in Table~\ref{TabPSF} because the data have been binned and rotated to sky coordinates which broadens the PSF (see below). Also, the fitting was performed using single Gaussians which does not represent the PSF shape so well. However, for this test we were looking for changes in the PSF width rather than measuring the PSF itself.

\begin{figure}
\begin{centering}
\includegraphics[width=60mm, angle=270]{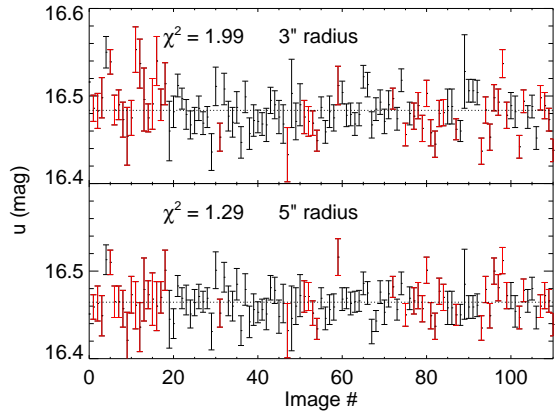}
\caption{The same star located at (03h32m55.62s, -27$^{\circ}$51\arcmin26.1\arcsec) in the E-CDF-S measured when the spacecraft is in sunshine (red points) and shadow (black points). Top: 3 arcsec aperture. Bottom: 5 arcsec aperture. Although the top plot has more scatter, it is uncorrelated with  the sunshine/shadow parameter described in Section~\ref{SecPSFVar}.}
\label{FigCountsPSF}
\end{centering}
\end{figure}

Despite the variability of the PSF, the photometry is not significantly affected if the standard  aperture of 5 arcecs is used.  This has been confirmed in several ways.  Assuming that the PSF varies uniformly it is expected to affect the measured flux by no more than 1.5 per cent. The photometry of sources observed in both states (sunshine and shadow) have been compared and no measurable effect is found when using a 5 arcsec aperture, either because the PSF distributions are so broad, or because there are other sources of scatter masking the effect. With a 3 arcsec aperture there is an increase in scatter, implying that the change in PSF is more significant closer to the core, but there is no evidence of a systematic change in photometry attributable to this PSF variation. This is illustrated in Fig.~\ref{FigCountsPSF} where repeated photometric measurements of a star in the  Extended Chandra Deep Field-South (E-CDF-S)  were made in sunshine and shadow, using the 3 arcsec and 5 arcsec aperture.

A smaller aperture is typically used for photometry of faint sources to inprove the signal to noise ratio. For a 3 arcsec aperture we suggest adding a systematic error of 0.015 magnitudes in quadrature to the random errors to account for the PSF variations. 

\begin{figure}
\begin{centering}
\includegraphics[angle=270,width=82mm]{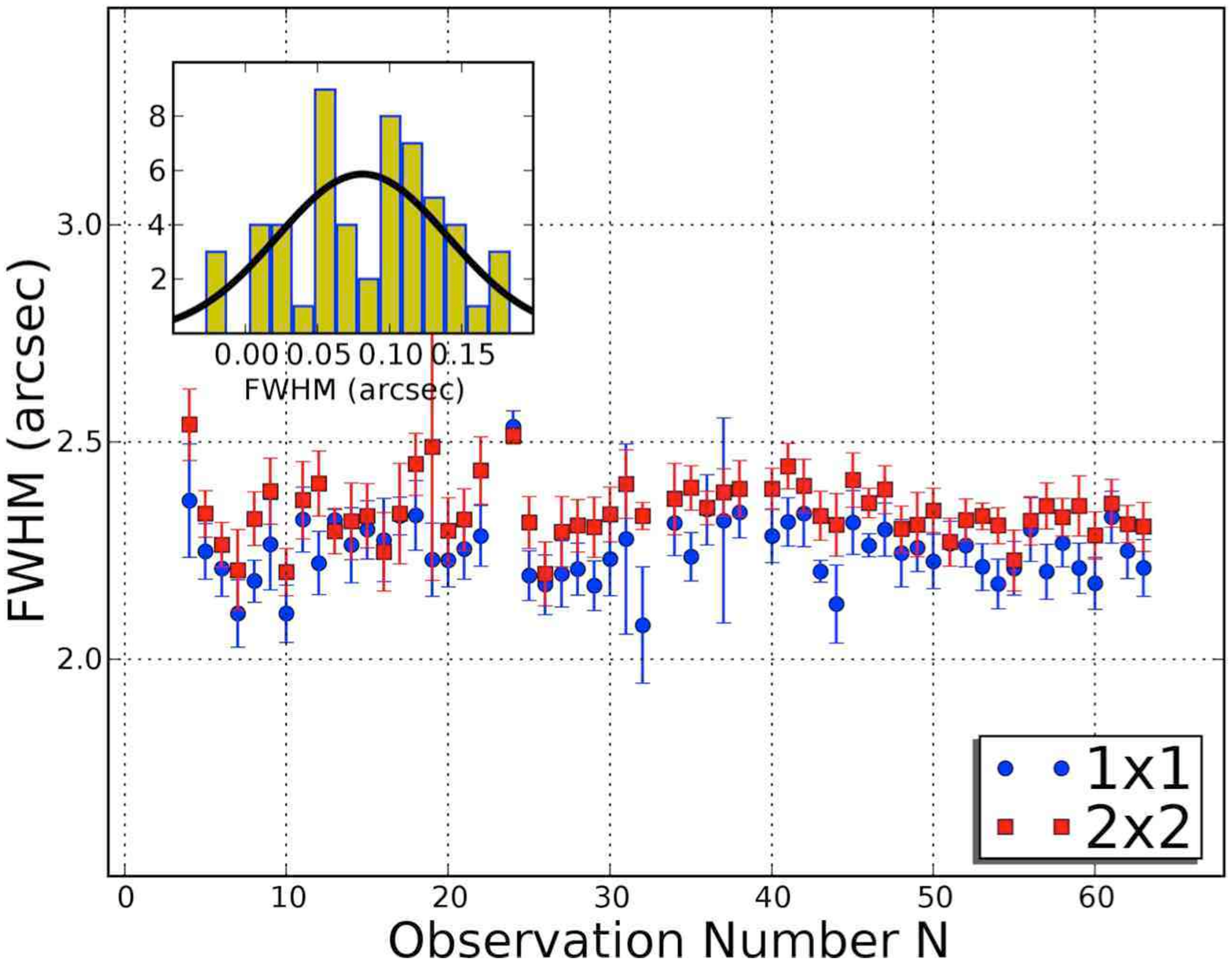}
\caption{The effect of binning on PSF width: the blue circles are measurements of the FWHM made from unbinned images, the red squares from the same images binned $2\times2$. The inset shows a histogram of the increase in FWHM width when images are binned from $1 \times 1$ to $2 \times 2$.}
\label{FigBinFwhm}
\end{centering}
\end{figure}

It is useful to note that the PSF also broadens slightly when an image is rotated (e.g. made into a sky image from a raw image), or binned (see Fig.~\ref{FigBinFwhm}). The effect of these combined can be as much as 0.15 arcsec, as shown in the inset in Fig.~\ref{FigBinFwhm}.  The core PSFs described here and recorded in the CalDB  were derived from raw, single exposure images and therefore this blurring is not included in the FWHM or core COGs. The PSF also varies slightly over the field of view, but the effect is less than 0.05 arcsec.  

\section{Coincidence loss in the high background regime}
\label{SecCoi}

Because UVOT is a photon counting detector it suffers from coincidence loss at high count rates \citep{FMG00}.  This is equivalent to ``pile-up'' in X-ray CCD detectors.  Coincidence loss occurs when two or more photons arrive at the same position within one CCD readout frame; only one photon will be counted, resulting in a systematic undercounting of the true photon flux. The error on the count rate is also affected; it is no longer purely Poissonian, but affected by the finite number of frames in an exposure \citep{Kuin}.  In order to make accurate measurements of photon fluxes with UVOT a coincidence correction needs to be made. 

The coincidence correction and error calculation described in Paper I, recorded in the CalDB and implemented in  \texttt{UVOTSOURCE} (included in \texttt{FTOOLS}) works well in most situations, however it still has several limitations:  it does not apply to crowded or extended sources and it was determined from observations with low background. However,  the background count rate is also a contributor to coincidence loss and the  background measured by UVOT depends on a number of factors.  In some cases, particularly with the {\it white} filter, or when the telescope is pointing close to the Earth or Moon, the  background can be high (see Section~\ref{SecBackground}).

\begin{figure}
\begin{centering}
\includegraphics[angle=0,width=84mm]{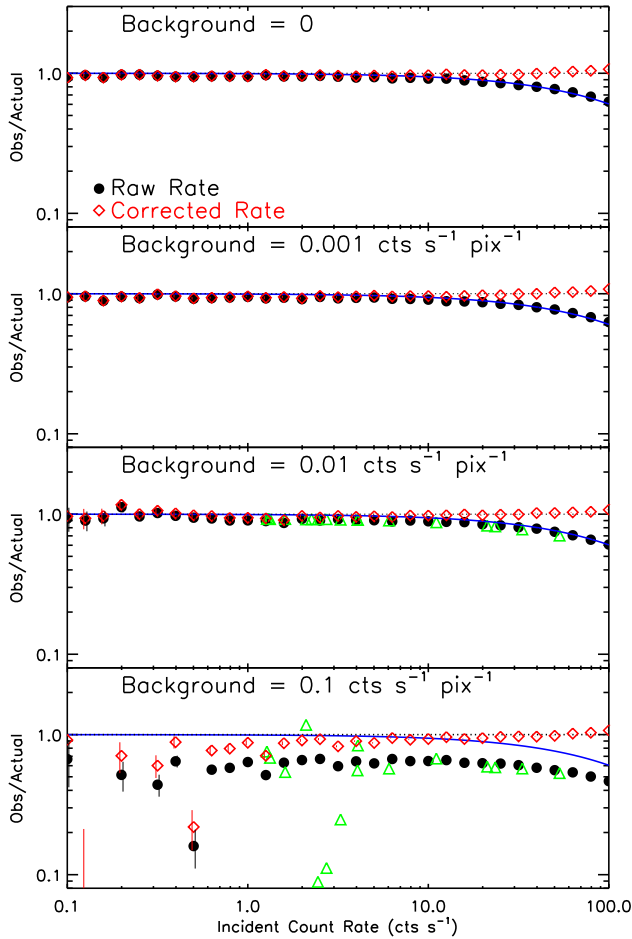}
\caption{High background coincidence simulation results.   Isolated point sources with a range of intensities and background  intensities are generated by the simulation and photometry is  performed using \texttt{UVOTSOURCE}.  The four panels show the ratio of the count rates measured by \texttt{UVOTSOURCE} to the true incident count rate generated by the model as a function of incident   count rate.  Filled black circles show the raw count rates measured,  and red diamonds show the count rates corrected for coincidence loss which is returned by \texttt{UVOTSOURCE}. Error bars  output by \texttt{UVOTSOURCE} are also plotted.  To compare the model predictions with real data, the green triangles show the raw count rate ratios measured from real UVOT data of a single field that has been observed with multiple background rates (see text in Section~\ref{SecCoi} for more details).  The blue line shows  the empirical coincidence loss model from Paper~I.  The top panel  shows the simulation with no background, while the second through  fourth panels have background count rates of 0.001, 0.01, and 0.1~s$^{-1}$~pix~$^{-1}$.}
\label{FigCoiPoint}
\end{centering}
\end{figure}

\begin{figure}
\begin{centering}
\includegraphics[angle=0,width=84mm]{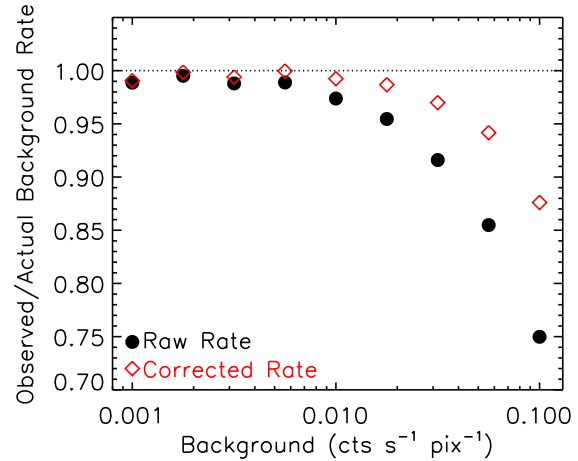}
\caption{The ratio of the measured to incident model background count  rate as a function of the incident background count rate for the  simulation shown in Fig.~\ref{FigCoiPoint}.  Filled black  circles show the raw background count rate and the red diamonds show   the coincidence-corrected rates measured by  \texttt{UVOTSOURCE}.  Similar to the case for the point sources in  Fig.~\ref{FigCoiPoint} the raw count rates for the background are underestimated to a larger degree as the background rate increases. However unlike the point sources the coincidence correction does not fully correct the background count rate.}
\label{FigCoiBack}
\end{centering}
\end{figure}

The sky is in some cases not the only source of background.  For example, UVOT carries out a program of observing supernovae in nearby galaxies \citep{Brown09}.  In this case the supernovae are point sources on top of a diffuse Galactic background which must be accurately subtracted as well as taken into account in the coincidence correction.  In other cases it is the background itself that is of scientific interest.  The ultraviolet surface brightness of galaxies can be converted into star formation rates.  Here the goal is to measure the surface brightness of a large, mostly uniform surface. Failure to correct for coincidence loss in this case will lead to an underestimation of the star formation rate and other measured values of the underlying stellar population. 

In order to determine the effects of high background on the coincidence-corrected photometry a model was constructed to simulate the UVOT instrumental behaviour.  This model was used to run simulations of UVOT photometry for point sources with count rates from 0.1 to 100~s$^{-1}$ and background count rates from 0 to 0.1~s$^{-1}$~pix$^{-1}$ (where a pixel is 0.502 arcsec; the real UVOT background count rate varies between 0 and 0.05~s$^{-1}$~pix$^{-1}$, and exceptionally can be as high as 0.35~s$^{-1}$~pix$^{-1}$, see Section~\ref{SecBackground}.)  Aperture photometry was then performed with \texttt{UVOTSOURCE} and the coincidence-corrected output was compared to the known source and background counts used as input to the model.

The first step of the model assumes a detector of 128 by 128 pixels and generates photons incident on the detector. The background photons are generated assuming a Poisson distribution with a mean equal to the background counts per frame.  Similarly the number of source photons per frame is selected from a Poisson distribution using the mean source counts per frame.  The source photons are then placed on the image of the frame using a 2 arcsec Gaussian point spread function.  Each frame image is then added to a master image which is a record of the incident photons without any UVOT instrumental effects.

Each of the `incident' frame images is passed through a model of the UVOT. Each photon is converted to a photon splash which is detected by the CCD.  The photon splash has a Lorentz profile with a full width half maximum of 24 microns or 1.09091 CCD pixels \citep{James07}.  The frame image is convolved with the photon splash to generate the image seen by the CCD.  The number of counts in each CCD pixel for each photon is measured and the centroiding performed, mimicking as closely as possible the onboard processing. 

It is necessary to vary the position of the centre of the source PSF positions relative to the CCD pixels. The reason for this can be understood by considering the two extreme cases: where the PSF is perfectly centred on a CCD pixel, and where the PSF is centred atop the vertex of four CCD pixels.  In the first case the bulk of the photons will fall on a single pixel, while in the second they will be split over four, with potentially different coincidence loss effects.  However, the pointing of \textit{Swift}  jitters with an amplitude of around 0.5 arcsec after the spacecraft settles.  A 0.5 arcsec positional jitter is therefore added in the model, which helps to dampen out the effect of the source positioning.

The model described above was run for a differing number of simulated frames for each pair of source and background flux values depending on the brightness of the source.  For the brightest sources ($>$10~s$^{-1}$) on the weakest backgrounds ($<$0.1~s$^{-1}$~pix$^{-1}$)  the model was run for 64\,000 frames (equivalent to an exposure time of 690~s), whereas for the weakest sources ($<$1~s$^{-1}$) on the highest background (0.1~s$^{-1}$~pix$^{-1}$) we used 2\,640\,000 frames or 29\,000~s.  The resulting model images were then analysed using \texttt{UVOTSOURCE} with a 5 arcsec aperture and the background region defined by an annulus.

%



The results of the simulation are shown in Figures~\ref{FigCoiPoint} and \ref{FigCoiBack}. Fig.~\ref{FigCoiPoint} gives the results for the photometry of the point sources.  Each panel shows the ratio of the measured count rate to the incident count rate as a function of the incident count rate in a 5 arcsec aperture.  The different panels have different background count rates ranging from no background at top to 0.1~s$^{-1}$~pix$^{-1}$ at bottom.  Filled black circles give the raw count rate measured by \texttt{UVOTSOURCE}.  Red diamonds show the count rate with the coincidence loss calculated by \texttt{UVOTSOURCE}.  Since the 5 arcsec aperture matches that used in the UVOT calibration there is no need for an aperture correction.  The blue line shows the empirical coincidence loss model from Paper~I.

The results of the simulation for point sources show that for low background levels the raw count rates match the empirical measurements from Paper~I.  For the high background case in the bottom panel the raw observed count rate experiences heavier coincidence loss than in the Paper~I measurement, as expected.  However all panels show that the corrected count rates from \texttt{UVOTSOURCE} recover the true count rates, at the 3 per cent level, over the background levels modelled, except in the case of a weak source and high background, where the large scatter is due to the poor signal to noise rather than a systematic error in the coincidence correction.  \texttt{UVOTSOURCE} photometry is generally robust over reasonable background levels.

To test the model we analysed a {\it white} field (containing the standard star WD1121+145) that has been observed with a range of background count rates. Twenty sources were measured in the low (0.01~s$^{-1}$~pix$^{-1}$) and high (0.1~s$^{-1}$~pix$^{-1}$) count rate regimes to find the differences between the raw and corrected count rates for the two cases. The measurements have been plotted as green triangles on Fig.~\ref{FigCoiPoint} for direct comparison. The `incident' count rates here are assumed to be the coincidence-corrected count rates from the lower background regime.  The higher background exposure lasted just 67s and therefore the fainter sources exhibit a lot of scatter. Nevertheless for the brighter sources the green triangles follow the curve given by the model confirming that the model is correctly predicting the UVOT behaviour.  

Fig.~\ref{FigCoiBack} shows the results from \texttt{UVOTSOURCE} photometry of the background itself.  In similar fashion to Fig.~\ref{FigCoiPoint}, Fig.~\ref{FigCoiBack} plots the ratio of the observed background count rate to the incident background rate as a function of the incident background rate.  Filled, black circles show the raw background count rate output by \texttt{UVOTSOURCE}, while red diamonds show the coincidence-corrected count rates.  Unlike the point source photometry, the corrected \texttt{UVOTSOURCE} background count rates are not fully corrected for coincidence loss.  However, Fig. \ref{FigCoiBack} reveals that for background count rates less than 0.01~s$^{-1}$~pix$^{-1}$ coincidence loss of the background can be disregarded with only a 1 per cent effect on the photometry.  This result is of particular use in the analysis of large, extended objects such as galaxies.  Provided that the surface brightness of such objects does not exceed 0.01~s$^{-1}$~pix$^{-1}$ they can be analysed using standard techniques with a maximum penalty of 1 per cent. In the future we intend to use the model to improve our coincidence correction of high backgrounds.

\section{Mod-8 noise}
\label{SecMod8}
As mentioned in the introduction, the final stage of the UVOT detector is a 256 by 256 pixel CCD. Individual events are centroided to one eighth of a physical CCD detector pixel by onboard electronics which are fast enough to operate in real time \citep{K94}. However the algorithm is intrinsically imperfect and leads to the subpixels having effectively slightly different sizes giving a modulation on an $8 \times 8$ grid \citep{Michel}, known as Mod-8 noise or Fixed Patterning.  An LED illumination is used to map this pattern and an onboard look-up-table re-distributes each photon according to the known pattern. However, some residual pattern remains in the images because of small gain variations over the face of the detector, and the simplicity of the onboard centroiding algorithm. 

\begin{figure}
\begin{center}
\leavevmode 
\psfig{angle=0,figure=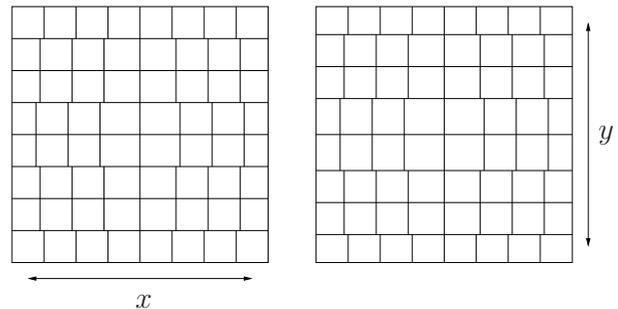,width=80mm}
\caption{Schematic of the pixel boundary determination and resampling in x and y in the mod-8 correction. In the left panel the pixel boundaries are adjusted in the x-direction to ensure that within any given row, there are equal counts per unit area in each pixel. This is followed by an adjustment of the row boundaries in the y-direction to give equal counts per unit area in each row.  The image is resampled using the new boundaries. This procedure is repeated starting with resampling in the y-direction followed by the x-direction, and the final mod-8 corrected image is made from the average of these two resamplings.}
\label{fig:ommod}
\end{center}
\end{figure}

The \texttt{UVOTMODMAP} tool, (included in \texttt{FTOOLS}), first steps a box of a chosen size across the image, and within the box a sigma clipping  algorithm is used to mask out sources of high significance and any surrounding pixels affected by coincidence loss. Then the average of all remaining mod-8 tiles within the sliding box is computed and used to produce a mod-8 map.

The mod-8 noise problem relates to the distribution of events, not to changes in the sensitivity, and so dividing the science image by the mod-8 map, whilst providing a simple cosmetic solution, would compromise the photometry of the image. Instead the algorithm resamples the image to give each image pixel equal area within a mod-8 tile. 

\begin{figure}
\begin{centering}
\includegraphics[angle=270,width=84mm]{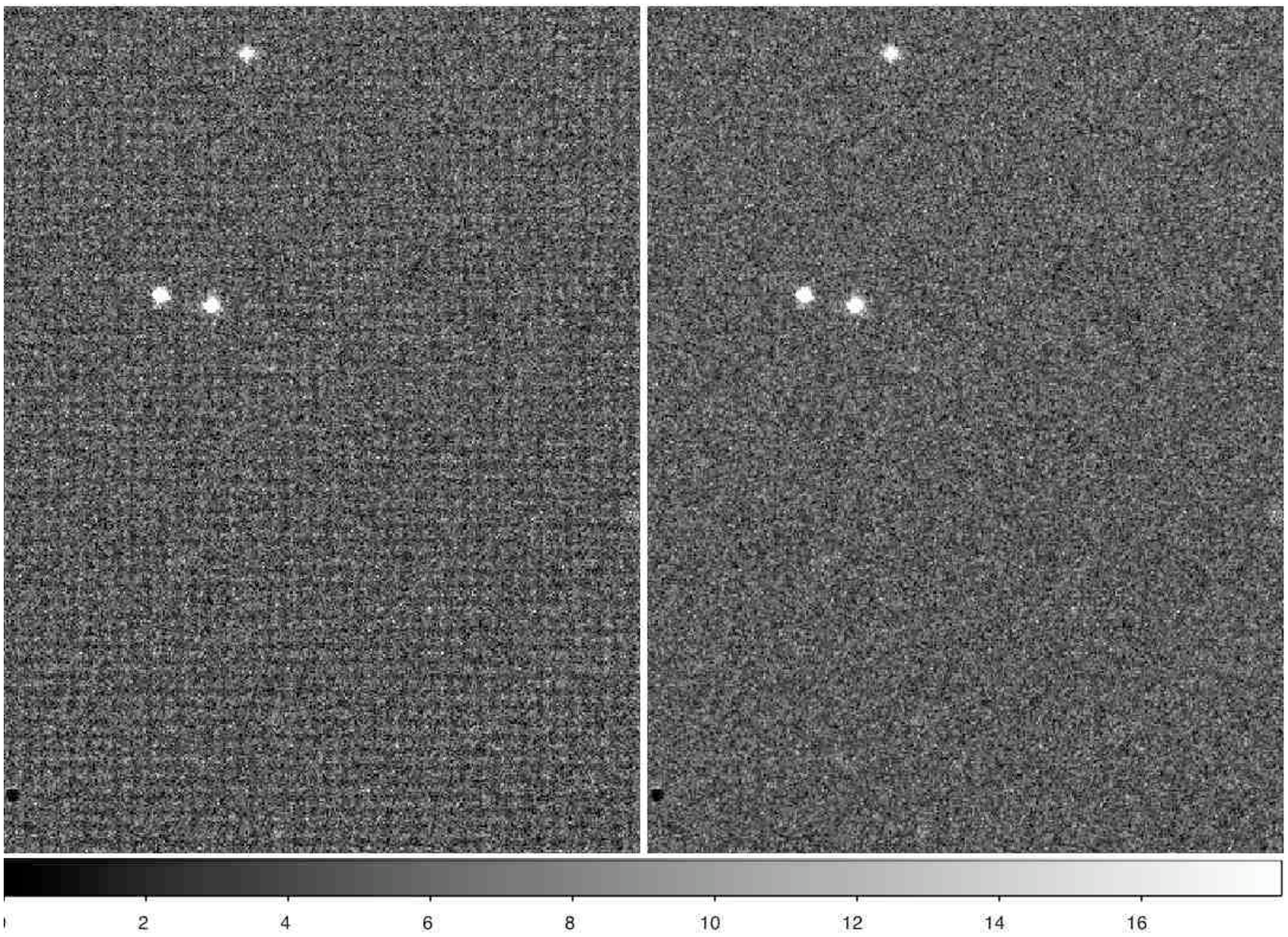}
\caption{Part of a raw {\it white} image before (left) and after (right) correction by \texttt{UVOTMODMAP}. The mod-8 patterning shows up as a faint grid in the left hand image. Each of these images is $470\times340$ pixels.}
\label{Figmod8}
\end{centering}
\end{figure}

For each mod-8 tile in the mod-8 map, the pixel x-boundaries are determined such that within each column of pixels the counts/unit pixel area is the same for each pixel, as illustrated in the left panel of Fig.~\ref{fig:ommod}. This is then taken as the spatial layout of the pixel x-boundaries in the mod-8 tile in the  science image and the science image is resampled to a grid of evenly spaced pixels. Then the y-boundaries of each row are determined such that each row has the same number of counts/unit pixel area, as illustrated in the right panel of Fig.~\ref{fig:ommod}. This is then taken as the spatial layout of pixel y-boundaries in the same mod-8 tile in the science image, which is then resampled to a grid of evenly spaced rows. 

The same procedure is then repeated with the order reversed (remapping in y followed by remapping in x) and the final corrected science image is made from an average of the two resamplings. This is to ensure that the redistribution of photons is performed in an identical fashion in x and y.

In Fig.~\ref{Figmod8}, a region of mostly sky background in a raw {\it white} image is shown before and after running \texttt{UVOTMODMAP}. The level of modulation before correction is about 7 per cent standard deviation and after correction it is reduced to 2 per cent. Because \texttt{UVOTMODMAP} is computationally intensive, it is not run in the standard pipeline for initial products, but it is run on the data before they are archived. 

It is important to note that the number of photons is conserved and therefore fixed patterning does not affect photometry, but it reduces the pixel-to-pixel scatter in the background, which means that the noise in the background is lower and thus faint sources stand out higher above the background noise.  The box-like pattern that appears around bright sources is not removed by this algorithm because it is due to coincidence loss.

\section{Large scale sensitivity}
\label{SecLSS}
The photon-counting nature of UVOT makes it insensitive to low-level CCD throughput variations, and so a traditional flatfield correction is not appropriate. However, UVOT photometry does show filter-dependent variations of up to 9 per cent with large-scale changes of position on the detector, presumably due to non-uniform sensitivity. We created a Large Scale Sensitivity (LSS) correction \citep{Landsman} by using repeated observations of stars at different positions on the detector.

The sensitivity variation is modelled as a 2-d quadratic with five free parameters:
\begin{equation}
LSS = 1 + c_1  x + c_2  x^2 + c_3  x  y + c_4  y + c_5  y^2
\label{eq:lss}
\end{equation}

where {\it x} and {\it y} are measured in pixels as distances from the centre (1024,1024) of the raw image.      The above formula ensures that the  LSS correction is unity at the centre of the image.       The coincidence-corrected count rate at a position must be divided by the LSS to yield the count rate that would be observed at the centre of the detector. 

\begin{figure}
\begin{centering}
\includegraphics[angle=0,width=84mm]{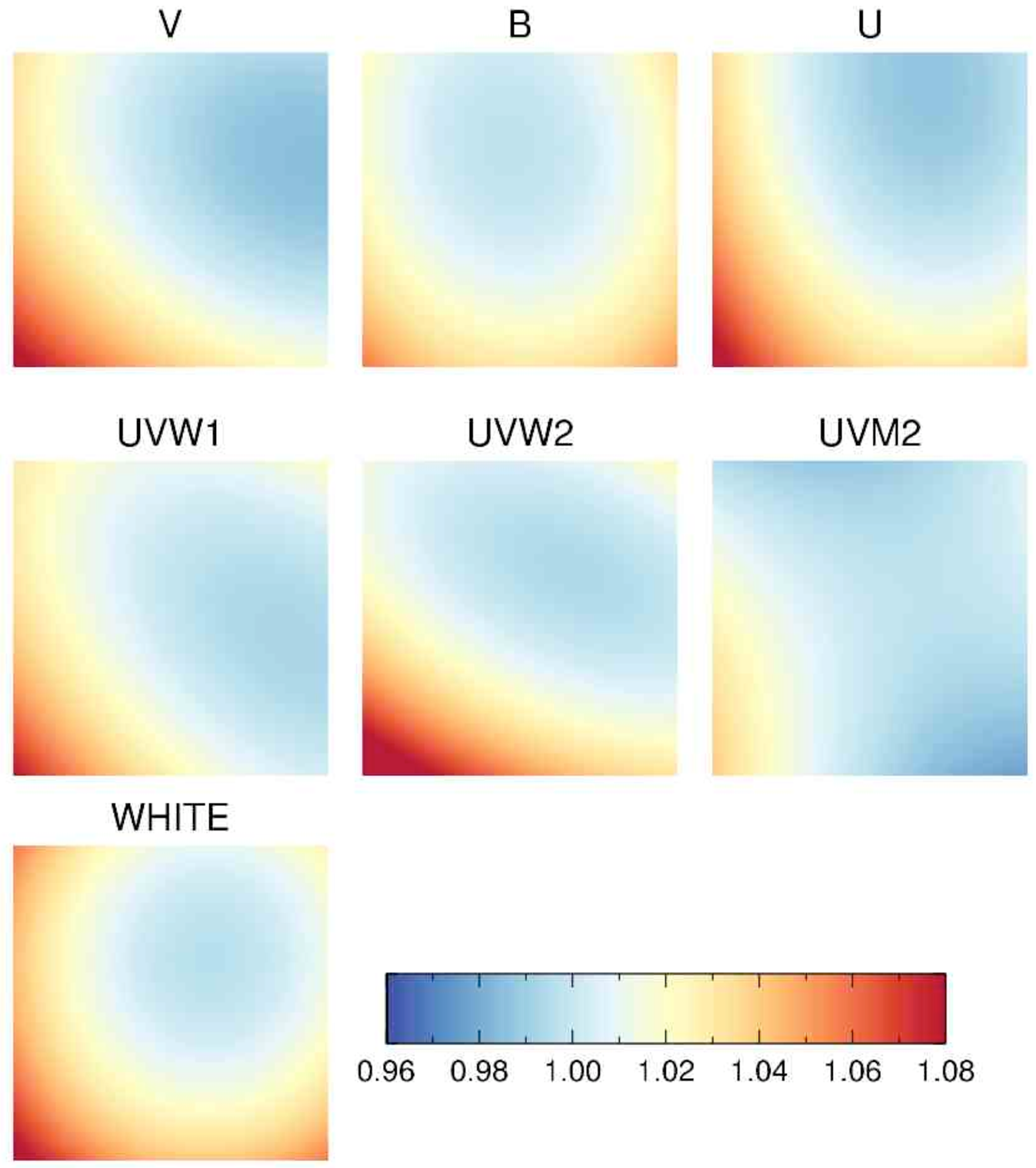}
\caption{ The correction due to large scale sensitivity variations for all the broadband filters as described in Section~\ref{SecLSS}. The colour indicates the magnitude of the correction. All filters show an increase in sensitivity in the bottom left hand corner, but the LSS for each filter is slightly different and cannot be used interchangably.}
\label{FigLSS}
\end{centering}
\end{figure}

To derive the coefficients, we used repeated observations of the same field, where the position of a star on the detector varied either because of explicit dithering or because the roll angle changed.  The fields include explicitly dithered observations of GD 50 ({\it v} filter) and NGC~188  ({\it white}), and long-term monitoring programs of 3C 279 ({\it u, b}) and the Galactic Centre ({\it uvw2, uvm2} and {\it uvw1}).     We  excluded variable stars, identified as those showing variability larger than the photometric errors despite minimal changes in the detector position.  We then used least-squares minimization techniques to adjust the parameters in Equation~\ref{eq:lss}  to minimize the variance in the stellar photometry.

Images of the LSS models are shown in Fig.~\ref{FigLSS}. The {\it v} map is similar to the coarse LSS sensitivity map shown in fig.~8 of Paper~I which was created using the same GD50 data.  All the filters are similar in having increased sensitivity toward the lower left-hand corner of the raw image. However, the LSS for each filter is distinct and not interchangeable. In particular the LSS correction for the {\it b} filter is smaller than for either the {\it u} or {\it v} filter and the correction for the {\it uvm2} filter is smaller than for the other two UV filters.

\begin{figure}
\begin{centering}
\includegraphics[angle=270,width=84mm]{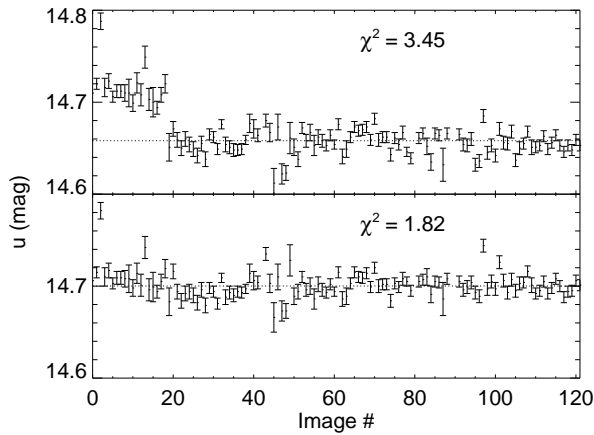}
\caption{ Repeated measurements of the same star located at (03h32m50.45s, -27$^{\circ}$48\arcmin33.0\arcsec) in the Chandra Deep Field: (above) with no LSS correction and (below) with LSS correction. The LSS clearly improves the {\it u} photometry for this star.}
\label{FigLSStest}
\end{centering}
\end{figure}

The use of the LSS improves the comparison between UVOT and external photometry in standard fields. For instance, UVOT photometry of stars in the PG 1633+099B field, with and without a LSS correction, has been compared with that of \citet{Stetson2000}. The use of the LSS significantly reduces the scatter between UVOT and Stetson photometry; for stars with $V<16.5$ the scatter is reduced from 0.033 mag to 0.024 mag. Fig.~\ref{FigLSStest} shows how the LSS also reduces the scatter in the UVOT {\it u} photometry in the case of a single star (located at 03h32m50.45s, -27$^{\circ}$48\arcmin33.0\arcsec in the CDF-S). This star was viewed 122 times over a 6 month period; the roll angle and raw detector position changed throughout that period. 

Note that the LSS correction must be applied {\it after} the non-linear coincidence correction.      For this reason, the LSS correction is not applied directly to the image, but rather is applied after the coincidence correction during a photometry calculation (i.e. using the LSSFILE option with the FTOOL program \texttt{UVOTSOURCE}).

\section{Small scale sensitivity}
\label{SecSSS}

\begin{figure}
\begin{centering}
\includegraphics[angle=270,width=84mm]{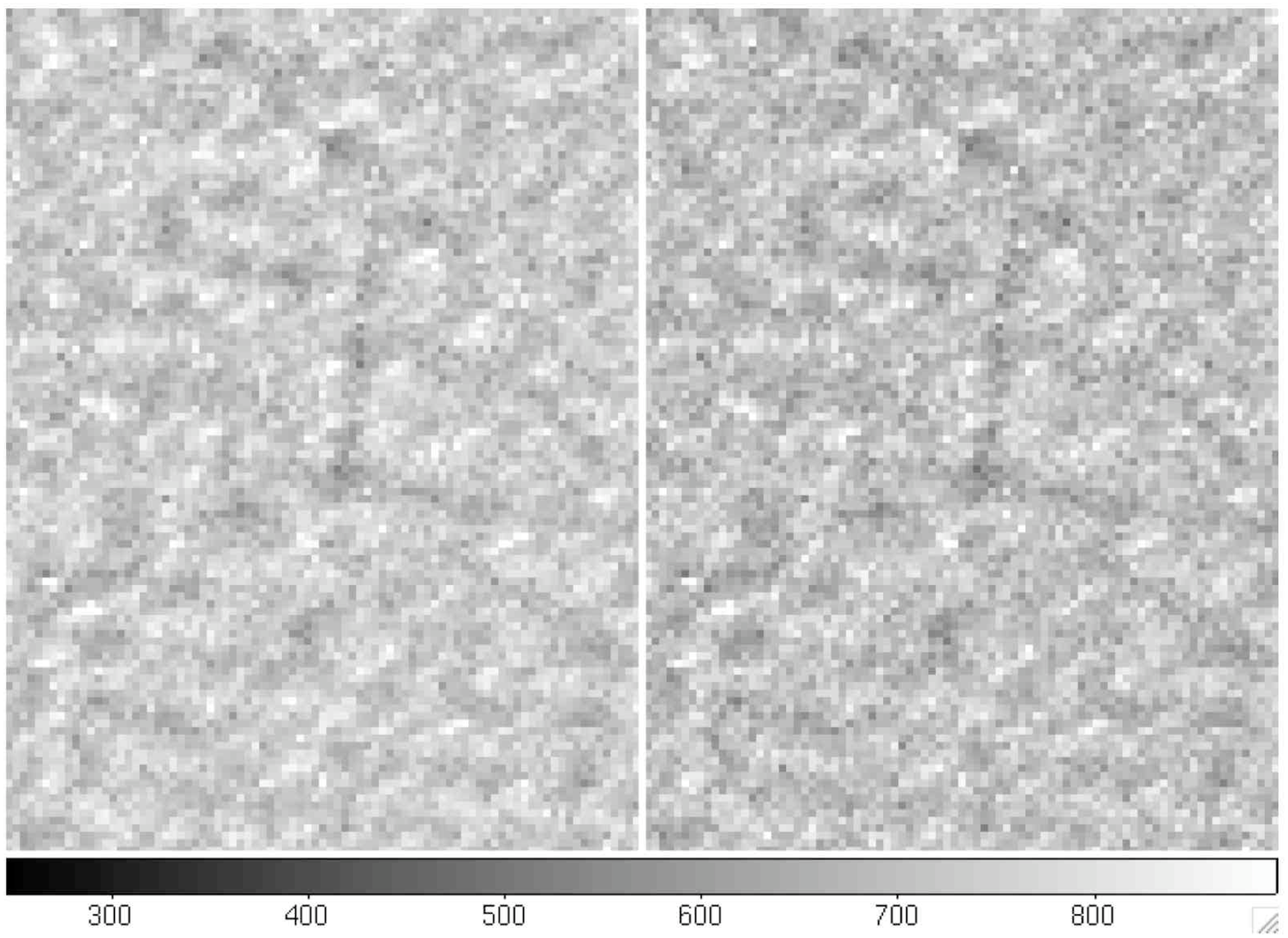}
\caption{Persistent small scale structure shows up in the UVOT flat fields, made using the onboard LED lamp. The SSS refers to the dark lines and patches over and above the regular mod8 noise. This is a comparison between data taken in 2005 (left) and 2008 (right). Each of these images is $120\times90$ pixels.  }
\label{FigCompimg}
\end{centering}
\end{figure}

The small scale variations in sensitivity (SSS) of the UVOT detector are measured by using an onboard LED lamp which illuminates the entire image fairly evenly; these images are known as `flat fields', and are smoothed over a large scale and summed up to make deep images.  There is a small scale structure visible in UVOT flat fields on the scale of a few pixels (see Fig.~\ref{FigCompimg}), most likely due to irregularities in the detector intensifier (MCP and fibre taper) or CCD. The standard deviation of counts per pixel in the centre of the image is about 
7 per cent if the mod-8 correction has not been performed (see Section~\ref{SecMod8}),  but if the image is binned over $8 \times 8$ pixels, (to the size of CCD pixels), then the variation falls to 2.4 per cent. 

This small scale structure seems to be stable with time, but because of the photon-counting nature of the detector it does not have a large effect on the count rate of a star summed in a 3--5 arcsec aperture. Repeated measurements of standard stars in different positions does reveal a variation in count rate higher than that predicted by Poisson statistics alone, but the variations are only slightly correlated with the SSS. For this reason we do not yet recommend any correction to the photometry for SSS. 

We might expect that if one area of the detector is used more heavily than other parts (e.g. the centre), we would see the sensitivity decreasing faster in this area.  However, comparing data taken in 2005 with those taken in 2008 shows no patches of decreasing sensitivity. 

\subsection{Bad pixels}

Some CCD pixels always produce very low or very high count rates and these are known as bad pixels and are  best removed from any analysis. To find the bad pixels, the same LED lamp exposures are used as for the SSS; the summed images are used to pick out low and high pixels. The count rates in most pixels lie in a Gaussian-type distribution about the average value.  Pixels which consistently have count rates more than three sigma from the mean are flagged as bad and recorded in the CalDB.

The number and location of bad pixels have changed very little since launch.  Apart from the pixels at the corners, and a wrap-around strip on the left hand side of the image, the only consistently 
bad pixels are listed in Table~\ref{TabBadPix}. 

\begin{table}
\caption{Bad pixels appearing in the centre part of the image. There are 3 groups of bad pixels: each group covers $8 \times 8$ pixels (corresponding to a CCD pixel). In this table we list the bottom left hand corner of each group.  }
\label{TabBadPix}
\begin{tabular}{@{}cc}
\hline
X (pixels) & Y (pixels) \\
\hline
31 &	719 \\
111 &	527 \\
1431 &	39 \\
\hline
\end{tabular}
\end{table}

\section{Astrometry}
\label{SecAstrometry}

\subsection{Distortion}
\label{SecDistortion}
The optical fibre taper in the detector intensifier introduces a positional distortion. Therefore the raw detector coordinates for a group of stars do not map linearly to  their relative positions on the sky. This can be rectified by applying a distortion correction.

\begin{figure}
\begin{centering}
\includegraphics[width=84mm]{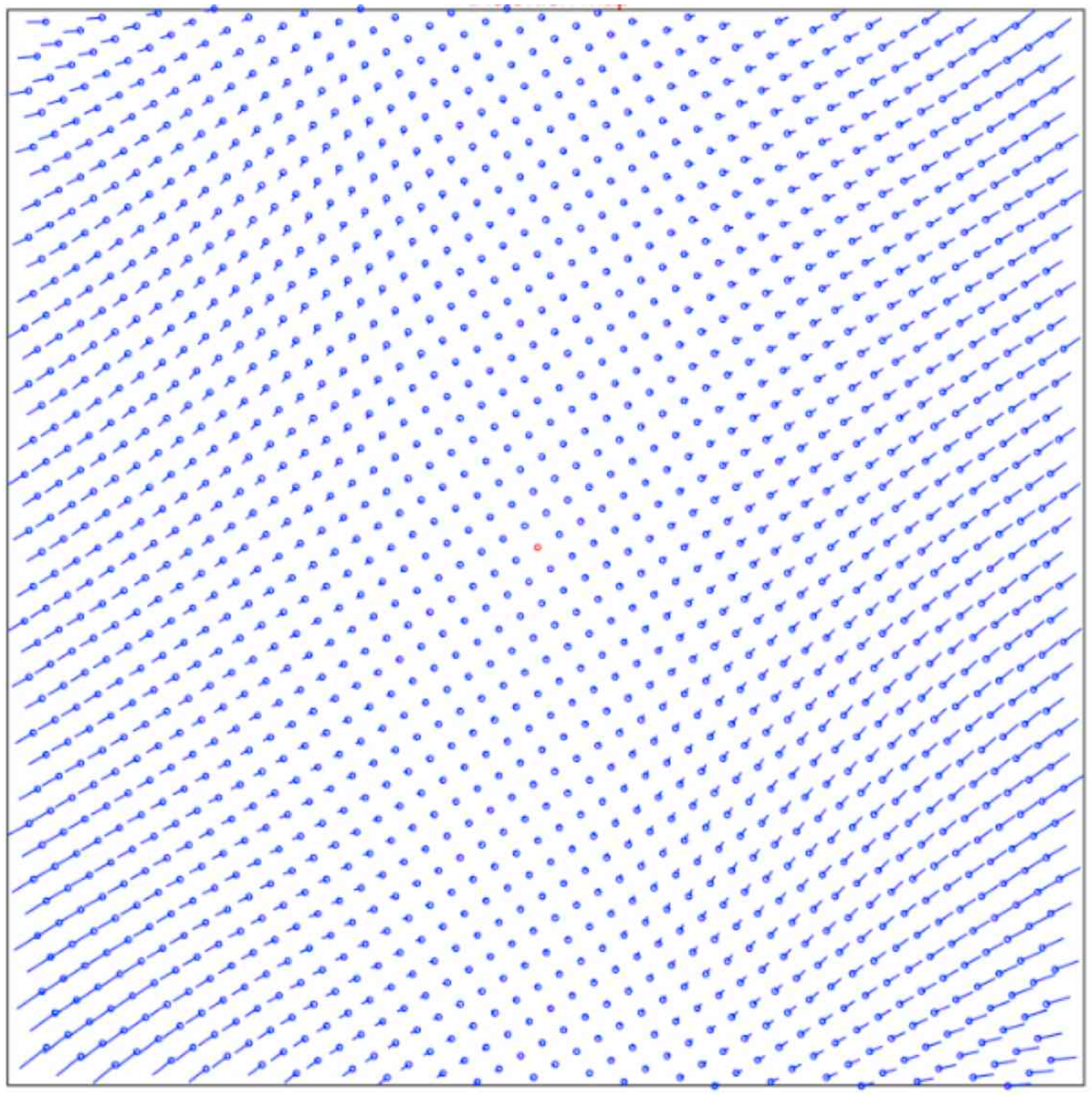}
\caption{The distortion map as described in Section~\ref{SecDistortion}. Each blue spot represents the position of the image of a pinhole in the calibration mask, with the line representing the shift required to undistort the grid of pinholes. }
\label{FigDist}
\end{centering}
\end{figure}

The distortion was mapped during the ground calibration using a target  mask with a regular grid of pinholes.    The ground-based distortion map was supplied as a set of 1952 correction vectors, with the size of the correction reaching 68 pixels near the edge of the detector.  The distortion is illustrated in Fig.~\ref{FigDist}, where the blue spots are at the measured positions of the pinholes, and the lines represent the vectors required to correct the image.  In the current UVOT pipeline, the distortion vectors are mapped on to a $256 \times 256$ grid using thin spline smoothing.  Two modifications were  made to the ground-based distortion corrections. The main change was to apply a rotation of  0.6$^{\circ}$ about the centre of the image to each displacement vector.   This change was initially suggested by comparison of over 1500 star positions  of a Magellanic Cloud target with source positions in the   Magellanic Cloud  Photometric Survey catalogue \citep{Zaritsky}.  A second small change to the ground-based distortion map was to delete two of the 1952 vectors  which had a large discrepancy with neighbouring vectors.   

\begin{figure}
\begin{centering}
\includegraphics[angle=270,width=70mm]{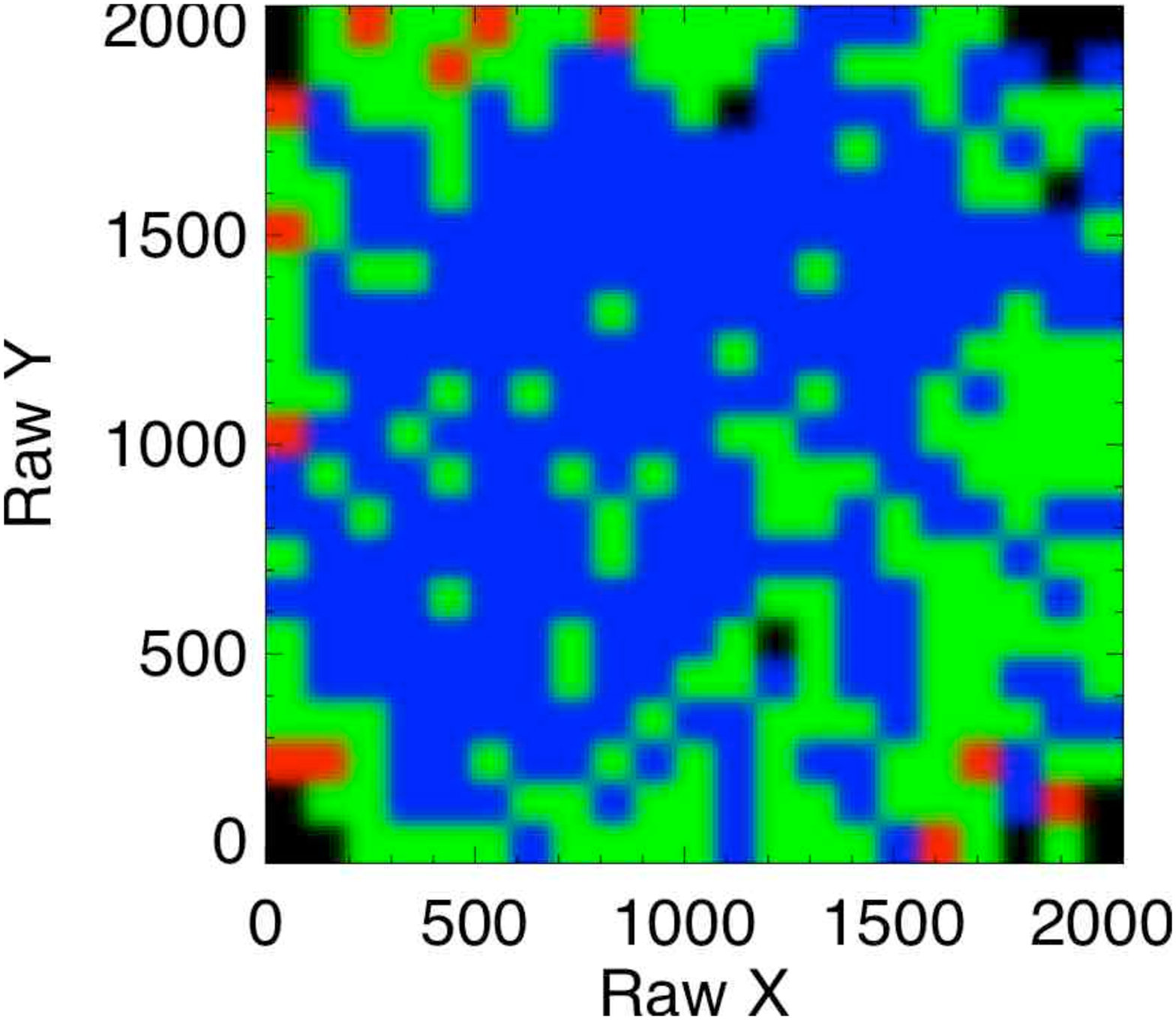}
\caption{The accuracy of the distortion correction. The $2048 \times 2048$ area of the UVOT CCD  is divided into a $21 \times 21$ grid.  Grid regions with a median astrometric error (compared to a high precision catalogue) of less than 0.2 arcsec are shown in blue,    regions with a deviation between 0.2 and 0.4 arcsec are shown in  green, regions with a median deviation greater than 0.4 arcsec are shown in red, and regions where no star positions have been measured are shown in black.}
\label{FigDistCorr}
\end{centering}
\end{figure}

Because of the distortion the raw UVOT image has a plate scale that varies between 0.47 arcsec~pix$^{-1}$ and 0.51 arcsec~pix$^{-1}$. The ground-based distortion map was intended to yield a uniform plate scale of 0.5 arcsec~pix$^{-1}$ but after correcting for distortion we find a slightly larger plate scale of 0.502 arcsec~pix$^{-1}$ for all filters except {\it uvw2}. The {\it uvw2} filter has a slightly larger plate scale of 0.504 arcsec~pix$^{-1}$ but is expanded in the UVOT pipeline to match the 0.502 arcsec~pix$^{-1}$ of the other filters.

The distortion is automatically removed from images when they are made into sky images in the pipeline, or by using \texttt{SWIFTXFORM} (included in \texttt{FTOOLS}) as a stand-alone tool. 

To test the accuracy of the distortion correction, we computed astrometric solutions for 30 images using high-precision star positions from  the Stripe 82 subset of the Sloan survey \citep{Ivezic}  and from the \citet{Stetson} catalogue of the open cluster NGC 188.      We then computed the deviation of the UVOT positions of 3107 stars on the 30 images from the catalogue positions.  In Fig.~\ref{FigDistCorr} the UVOT imaging area is divided into a $21 \times 21$ grid, and a color code is supplied for the median astrometric deviation of the stars within each grid area.    Over most of the detector the median deviation is under 0.2 arcsec, though there is significant increase toward the detector edges.     The poorer astrometry near the detector edge is probably  due to  the larger distortion correction there (see Fig.~\ref{FigDist}), but a degradation of the PSF near the detector edge might also contribute.

\subsection{Boresight and aspect correction}
\label{SecBoresight}

The UVOT boresight is defined as the average position on the UVOT detector where the targeted source is found. There are variations in the position because of the pointing (or attitude) knowledge of the {\it Swift} attitude control system. We constructed distributions of the offsets from the nominal boresight in x- and y-detector coordinates. To determine the true boresight values we fitted Gaussians to the distributions. The (1-sigma) width of the Gaussians is typically about 1.3 arcsec, and this is a reasonable estimate of the pointing knowledge of the {\it Swift} pointing system.

\begin{figure}
\begin{centering}
\includegraphics[angle=270,width=84mm]{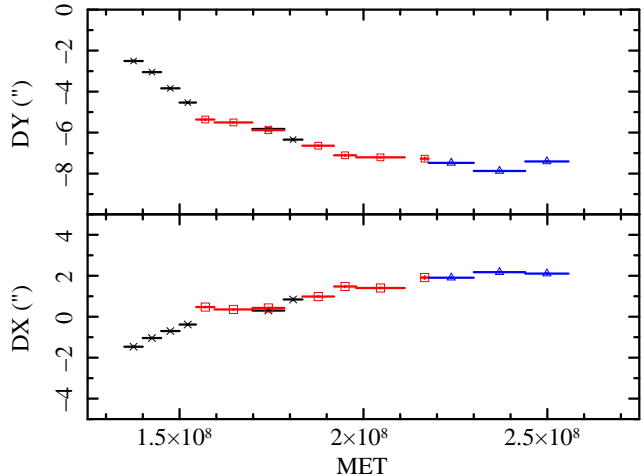}
\caption{ The history of the {\it v} filter boresight. Each colour and symbol indicates the telescope definition (teldef) file used to process the data and calculate the offset. The values have been adjusted to account for the changing teldef files so that they provide the actual changes to the boresight.}
\label{FigHistBS}
\end{centering}
\end{figure}

The boresight position is found to depend on the filter used in the observation, and it also changes with time at a rate of approximately 1 arcsec per year. This effect is due to a drift between the UVOT boresight and the satellite star tracker boresight and a similar drift is also seen in the XRT \citep{Moretti}. Fig.~\ref{FigHistBS} shows the gradual change with time for observations using the {\it v} filter. The evolution for the other UVOT filters is similar to that of the {\it v} filter. 

The boresight values are used to generate the telescope definition (teldef) files in the CalDB, and the values themselves are documented in the teldef files. Additional teldef files were added to CalDB in May 2009, to track the evolution of the boresight positions. In Fig.~\ref{FigHistBS} the different colours represent boresights measured with different teldef files. The teldef files are used in the standard {\it Swift} processing pipeline to create sky images and determine sky coordinates for events. 

The attitude solution provided by the spacecraft attitude control system and knowledge of the boresight can, in the vast majority of images, be improved by matching sources detected with UVOT to entries in a star catalogue. The {\it Swift} pipeline detects sources using \texttt{UVOTDETECT} (included in \texttt{FTOOLS}) which calls \texttt{SEXTRACTOR} \citep{Bertin}, and then matches sources with the USNO-B1 catalogue \citep{USNO-B}. Using the matched sources, the software determines the best rotation in a least squares sense to align the positions of the detected sources with the positions of the stars in the catalogue. The RA and Dec of the centre of the image are then adjusted. The measured offset in roll about the target position is typically very small ($\sim1$ arcmin), and is not used in the aspect correction. In some cases, especially if the field is crowded, the automatic aspect correction can fail. In these cases it is sometimes possible to perform aspect correction by supplying an alternative reference catalogue to the tool \texttt{UVOTSKYCORR}.

\subsection{Astrometric accuracy}

\begin{figure}
\begin{centering}
\includegraphics[angle=0,width=70mm]{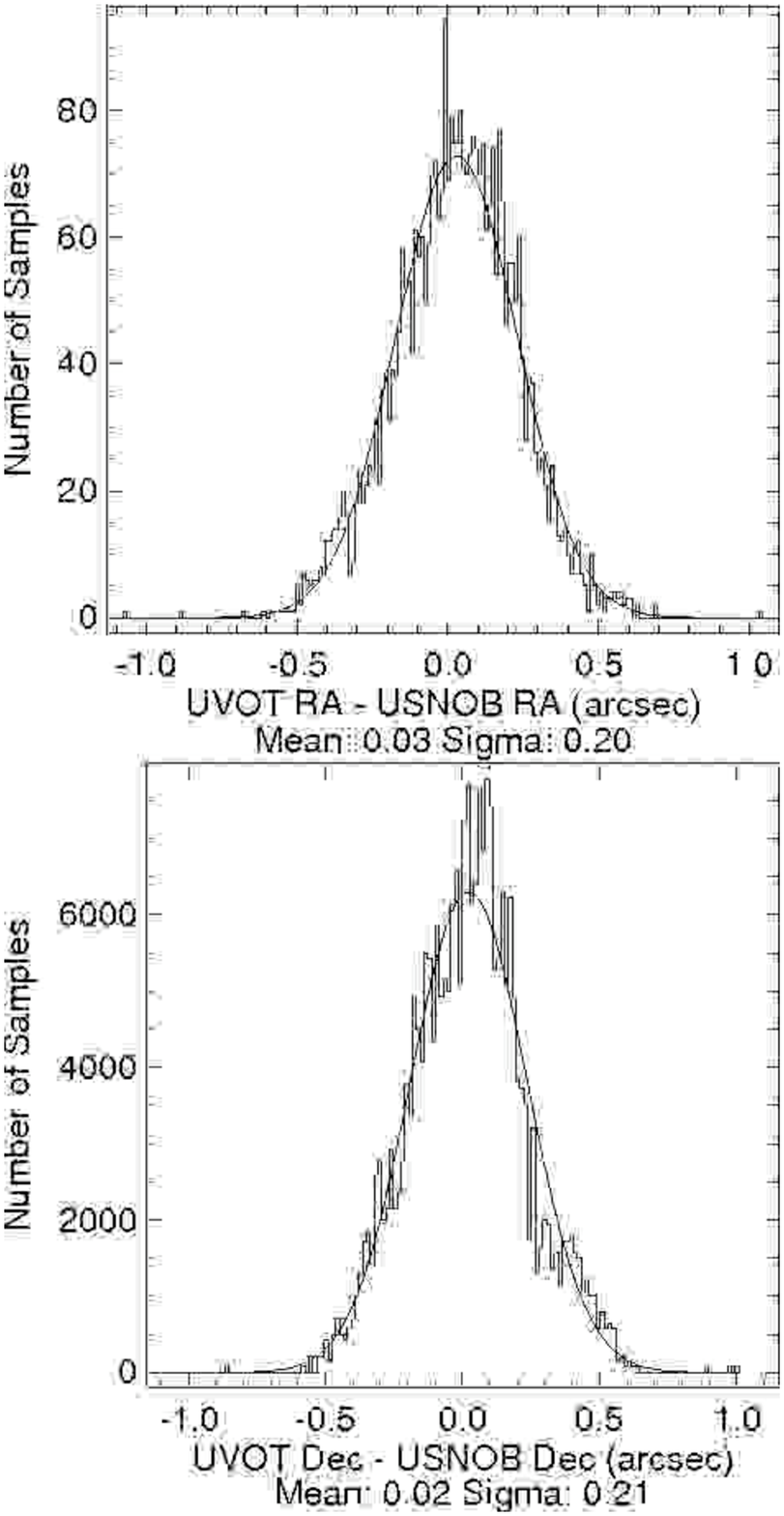}
\caption{ Comparing star positions as measured by UVOT versus those measured by USNO-B1. The top panel compares coordinates in RA; the bottom in Dec. The mean offset and distribution widths are given beneath each plot.}
\label{FigUSNO}
\end{centering}
\end{figure}

\begin{figure}
\begin{centering}
\includegraphics[angle=0,width=60mm]{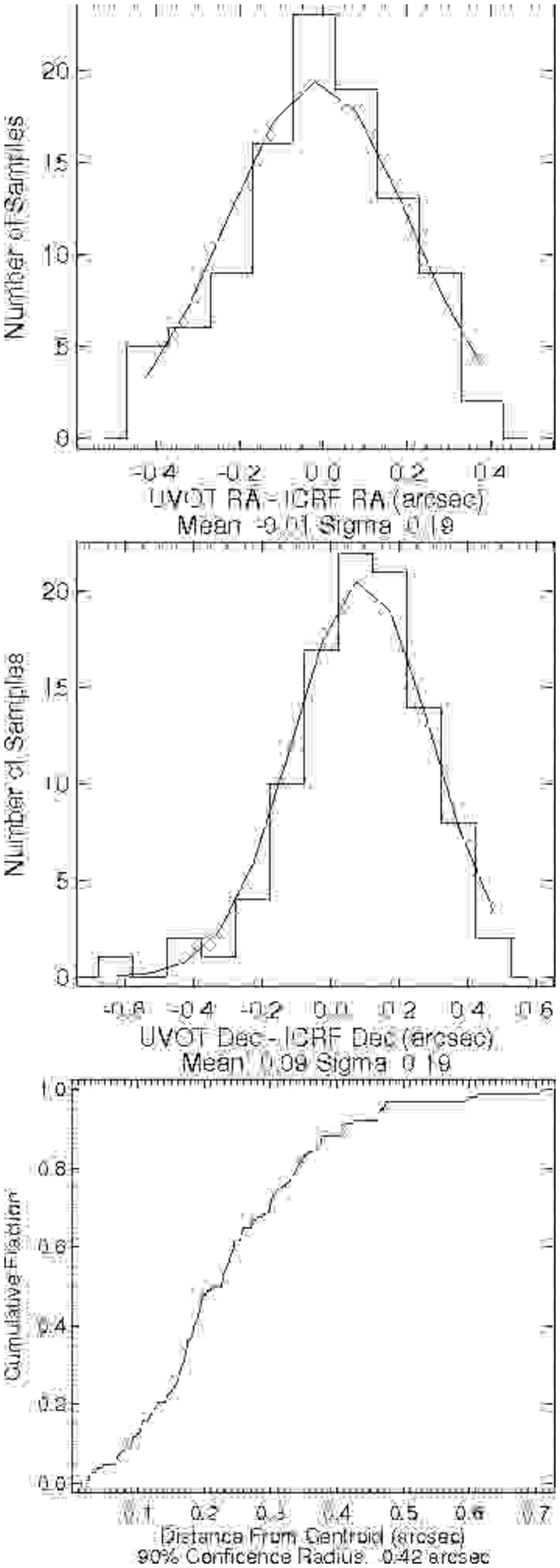}
\caption{ Comparing star positions as measured by UVOT versus those given in the ICRF. The top and middle plots show the offset in RA and Dec respectively from the ICRF coordinates. The mean offset and distribution widths are given beneath each plot. The bottom plot shows the positional confidence radius when compared to ICRF}
\label{FigICRF}
\end{centering}
\end{figure}

The self-consistency of UVOT astrometric solutions has been measured by comparing UVOT source positions after aspect correction, to USNO-B1 positions (cf. \citealt{Roming09}).  We have also measured UVOT's absolute astrometric accuracy by comparing UVOT source positions to VLBI-derived positions of high redshift quasars in the International Celestial Reference System \citep{ICRS}.  

For our self-consistency test, we chose as our target population field stars from {\it Swift} GRB observations.  The reasons to use GRB fields are: GRB fields are randomly distributed on the sky and  they also tend to have several kiloseconds of data in multiple lenticular filters.

We identified in the SDSS catalogue 108 point sources in 32 UVOT GRB fields that were within 2 arcmin of the burst. Images with contaminating features such as charge trails, bad aspect solutions or diffraction spikes were discarded from our test set. Likewise, sources brighter than 15th magnitude and fainter than 19th magnitude were rejected to reduce the effects of coincidence loss or background noise. Finally, we rejected sources detected at lower than a 10-sigma level and those within the halo of neary bright stars, again to reduce the effect of background noise on our results.  We began by using only unbinned images in our analysis.  After data selection, 3368 astrometric samples remained in our population.  This represented a total of 83 sources in 28 GRB fields. 

Photometric measurements were performed on each source to allow us to discriminate by magnitude and  \texttt{UVOTDETECT}  was then run on each image.  The result was a database of photometric and astrometric measurements.

Fig.~\ref{FigUSNO}  shows a histogram of the offsets between UVOT positions and those in USNO-B1 for the same objects, in RA and Dec.  There is good agreement between UVOT sky images and USNO-B1 (see the first two lines in  Table~\ref{TabAstrometry}).   To check the \texttt{UVOTDETECT} positions, we used an alternative method of determining the source positions by fitting a 2-D Gaussian to each well-detected source using IDL.  The results agree nicely with the results shown here.

Two fields in our population had sufficient data in multiple filters to investigate filter dependence of UVOT aspect solutions:  we see no differences between the filters.  We also investigated the effect on the astrometry of the lower resolution of binned images.  For 10 of the GRB fields in our sample, data were available in both binned and unbinned images. We compared the positions of the sources in these fields and found that binning degrades the astrometry by less than 0.03 arcsec in both directions. 

Our conclusion is that the astrometry on UVOT sky images is well calibrated to the USNO-B1 astrometric system in all 7 UVOT filters.  

The USNO-B1 catalogue is based on the FK5 celestial reference system \citep{Fricke1988}, but in 1997 the IAU adopted the ICRS \citep{Feissel1998} as the new reference system which is based on the VLBI positions of extragalactic sources. The extragalactic sources have fixed positions which eliminates the effects of proper motion. The ICRS Product Center\footnote{http://hpiers.obspm.fr/icrs-pc/} provides positions for a number of radio sources in the International Celestial Reference Frame (ICRF).  Their database gives precise positions (typically less than 0.002 arcsec per axis), redshifts, V magnitudes, and object types. We found 104 high redshift quasars with UVOT detections.

The upper two panels of Fig.~\ref{FigICRF}  show a comparison, in RA and Dec respectively, of UVOT derived and ICRF positions of our target population.  There is nice agreement between ICRF and UVOT in RA, but we found a systematic offset of 0.09 arcsec in Dec between ICRF and UVOT (see Table~\ref{TabAstrometry}).  These offsets are consistent in magnitude and direction with the expected difference between the two reference frames \citep{Mignard2000}, 
plus an average proper motion of 0.0025 arcsec per year.  We measure an absolute astrometric accuracy 90 per cent confidence interval of $0.42 \pm 0.03$ arcsec, which is consistent with earlier measurements. This systematic error can be added in quadrature with the statistical position error.

\begin{table}
\caption{Systematic and Random errors in the UVOT source positions as compared with the same sources in USNO-B1.0 and the International Celestial Reference System. The USNO-B1 comparison demonstrates that the UVOT astrometry is well calibrated. The ICRS result gives a measure of the absolute accuracy.}
\label{TabAstrometry}
\begin{tabular}{@{}ccc}
\hline
Error (arcsec):    & Systematic & Random \\
\hline
RA(UVOT)-RA(USNO-B1) & $0.03 \pm 0.003$ & 0.20 \\
Dec(UVOT)-Dec(USNO-B1) & $0.02 \pm 0.004$ & 0.21 \\
RA(UVOT)-RA(ICRF) & $-0.01 \pm 0.002$ & 0.19 \\
Dec(UVOT)-Dec(ICRF) & $0.09 \pm 0.002$ & 0.19 \\
\hline
\end{tabular}
\end{table}

\section{Background}
\label{SecBackground}
There are several sources of background counts which have to be subtracted to get true source count rates. The most obvious is the general sky background which comes from stellar photons scattered by the interstellar medium and solar photons scattered within our solar system and the Earth's atmosphere,  but we also have contributions from scattered light inside the telescope and detector system, as well as the low-level count rate (dark current) from the detector itself.

\subsection{Background statistics}

We have measured the background on randomly selected raw images, from which sources have been removed using a sigma-clipping algorithm. Images with exposures shorter than 100s or containing large extended objects were excluded. Table~\ref{TabSky} gives the average number of background counts measured in a 5 arcsec radius aperture, along with the 10th and 90th centiles to give an idea of the spread in values.  
\begin{table}
\caption{The background count rate in a 5 arcec aperture (314 pixels). The second column gives the number (N) of images used in the measurement. The next 3 columns give the background value for the 50th, 10th and 90th centile in the distribution. }
\label{TabSky}
\begin{tabular}{@{}lcccc}
\hline
Filter & N & 50\% & 10\% & 90\% \\
\hline
{\it v} &        12654 &    2.12 &    1.16 & 3.90 \\
{\it b} &          5820 &    3.36 &    1.78 & 6.68 \\
{\it u} &        12306 &    1.62 &    0.79 & 3.10 \\
{\it uvw1} & 18564 & 0.238 & 0.126 & 0.474 \\
{\it uvm2} & 10685 & 0.060 & 0.041 & 0.122\\
{\it uvw2} &  13688 & 0.114 & 0.068 & 0.264 \\
{\it white} &    3626 &    8.11 &   4.55 & 14.03 \\
\hline
\end{tabular}
\end{table}

\begin{figure}
\psfig{figure=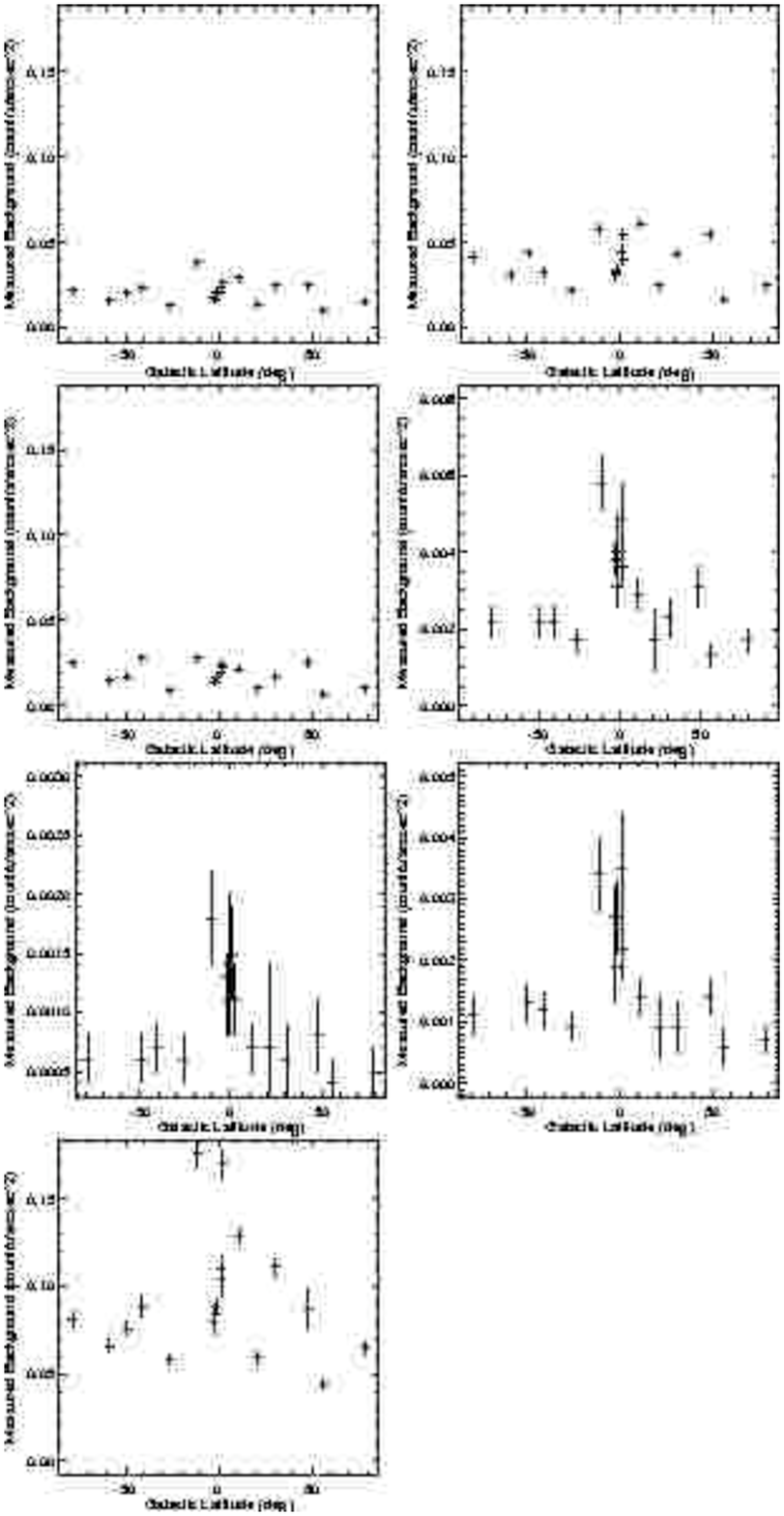, angle=0,width=84mm}
\caption{The measured background in all filters plotted with respect to Galactic latitude. From left to right, top to bottom the filters are: {\it v, b, u, uvw1, uvm2, uvw2, white}. }
\label{FigAllsky}
\end{figure}

\begin{figure}
\begin{centering}
\includegraphics[angle=270,width=84mm]{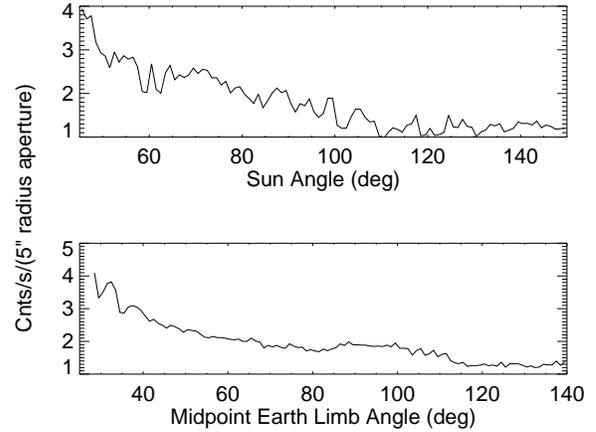}
\caption{The background count rate in all filters is found to vary with the angle to the Sun or the Earth. In this plot the background is measured in a 5 arcsec aperture in the {\it v} filter and is found to more than double as the angle towards the Sun (top) or Earth (bottom) reduces. }
\label{FigVsky}
\end{centering}
\end{figure}

\begin{figure}
\begin{centering}
\includegraphics[width=84mm]{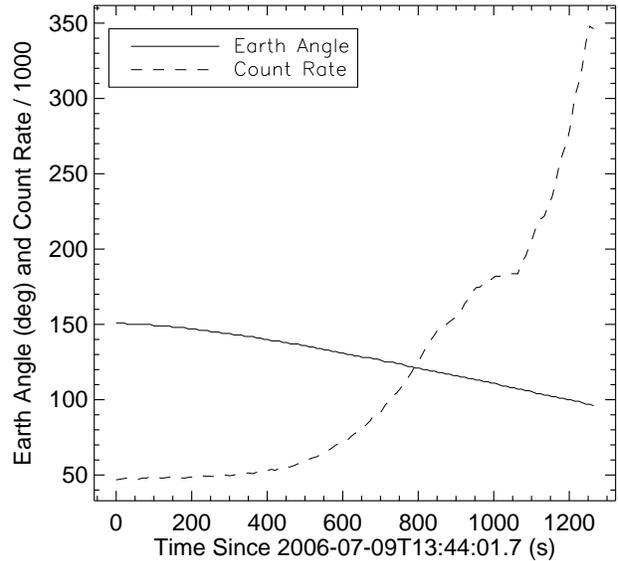}
\caption{The {\it white} filter is most sensitive to the changing Earth limb angle. This plot shows a time series taken from event mode data taken with the {\it white} filter as the Earth angle changes. The background count rate over the whole detecgtor increases from 50,000 to 350,000~s$^{-1}$ as the Earth limb angle changes from 150 to 100 degrees. This is the most extreme case so far seen. Above 192,000~s$^{-1}$ events are discarded by the camera leaving part of the image blank.}
\label{FigPhilwhite}
\end{centering}
\end{figure}

The actual background measured in an individual image depends on a number of factors apart from the filter, including the Galactic latitude, the ecliptic latitude and the Earth limb angle. To measure the effect of the Galactic latitude we searched the UVOT data archive for observations with at least 1000 seconds in all optical and UV filters and selected a number of fields at various Galactic latitudes.  We summed the images in each filter, histogrammed the count rates in each pixel, and fitted them with a Gaussian profile.  No attempt was made here to filter out the effects of bright Earth shine (see below).  

Fig.~\ref{FigAllsky} shows the relationship between the measured backgrounds and the Galactic latitude for each filter. The UV data show a marked dependence on the Galactic latitude; the optical less so. The sky background was predicted pre-launch for each filter using a model of zodiacal and Galactic light recorded in the CalDB (\citealt{Poolea} and \citealt{Pooleb}). Comparing these measurements with our predictions we find fairly good agreement (within a factor of 3) in all filters, even though the model does not include scattered light or dark current. 

The bright Earth can significantly increase the background of individual images as the angle of the line of sight to the Earth reduces.  Fig.~\ref{FigVsky} shows an example of this for the {\it v} filter and  Fig.~\ref{FigPhilwhite} shows the most extreme case seen so far in the {\it white} filter. To minimize the effect of the bright Earth limb the UVOT can be used in ``Earth limb protection mode'', where the exposure is broken into multiple exposures, with short exposures when  {\it Swift's} pointing will be near the Earth's limb, allowing high background exposures to be discarded if necessary.

\subsection{Scattered light}

\begin{figure}
\begin{centering}
\includegraphics[width=84mm]{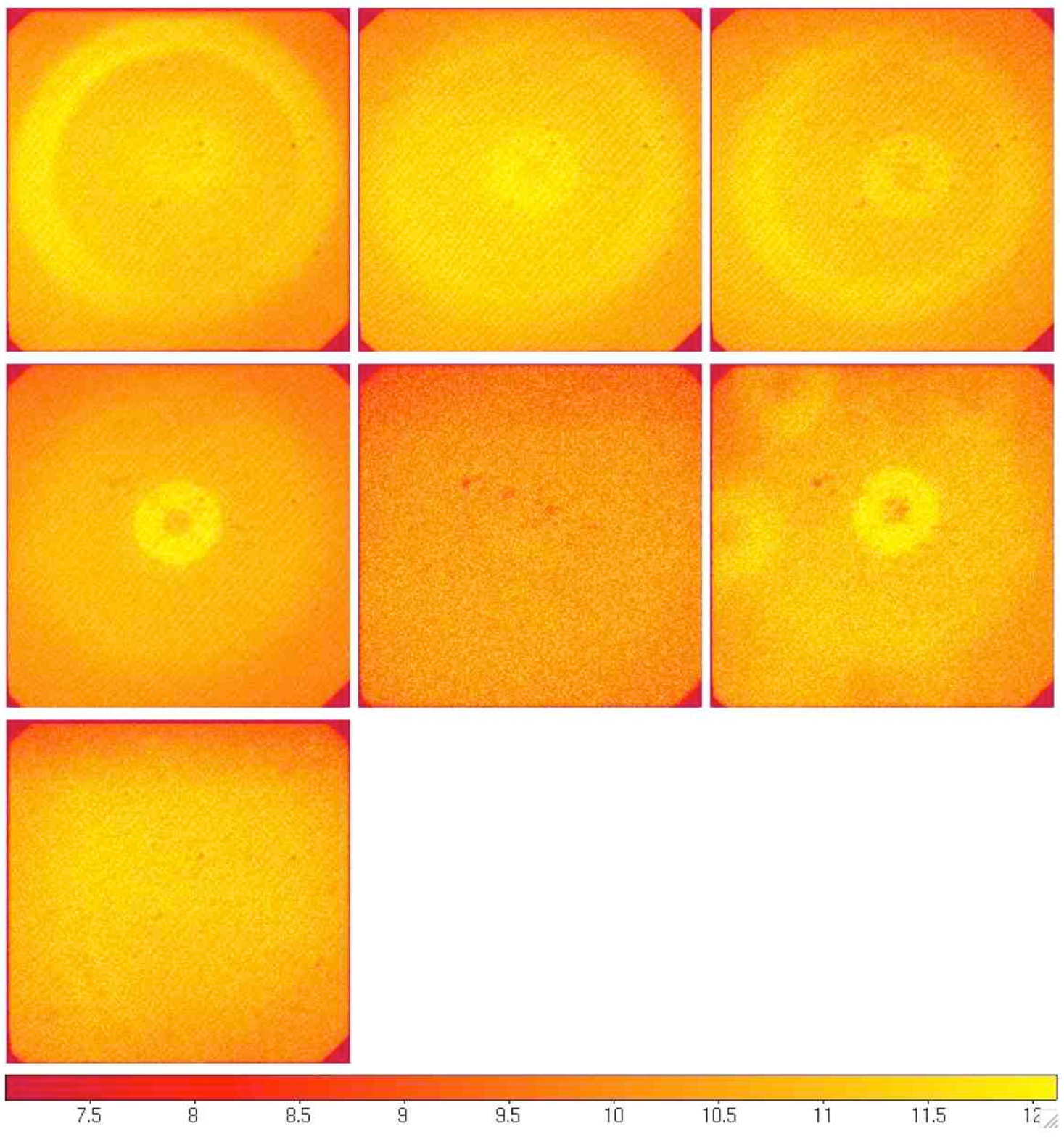}
\caption{The normalised scattered light images in top row: {\it v, b, u}, middle row: {\it uvw1, uvm2, uvw2} and bottom row: {\it white}. The scaling has been chosen to enhance the scattered light and is set the same for each panel. The {\it white} and {\it uvm2} images do not show marked rings.} 
\label{FigScat}
\end{centering}
\end{figure}

\begin{figure}
\begin{centering}
\includegraphics[width=80mm,angle=270]{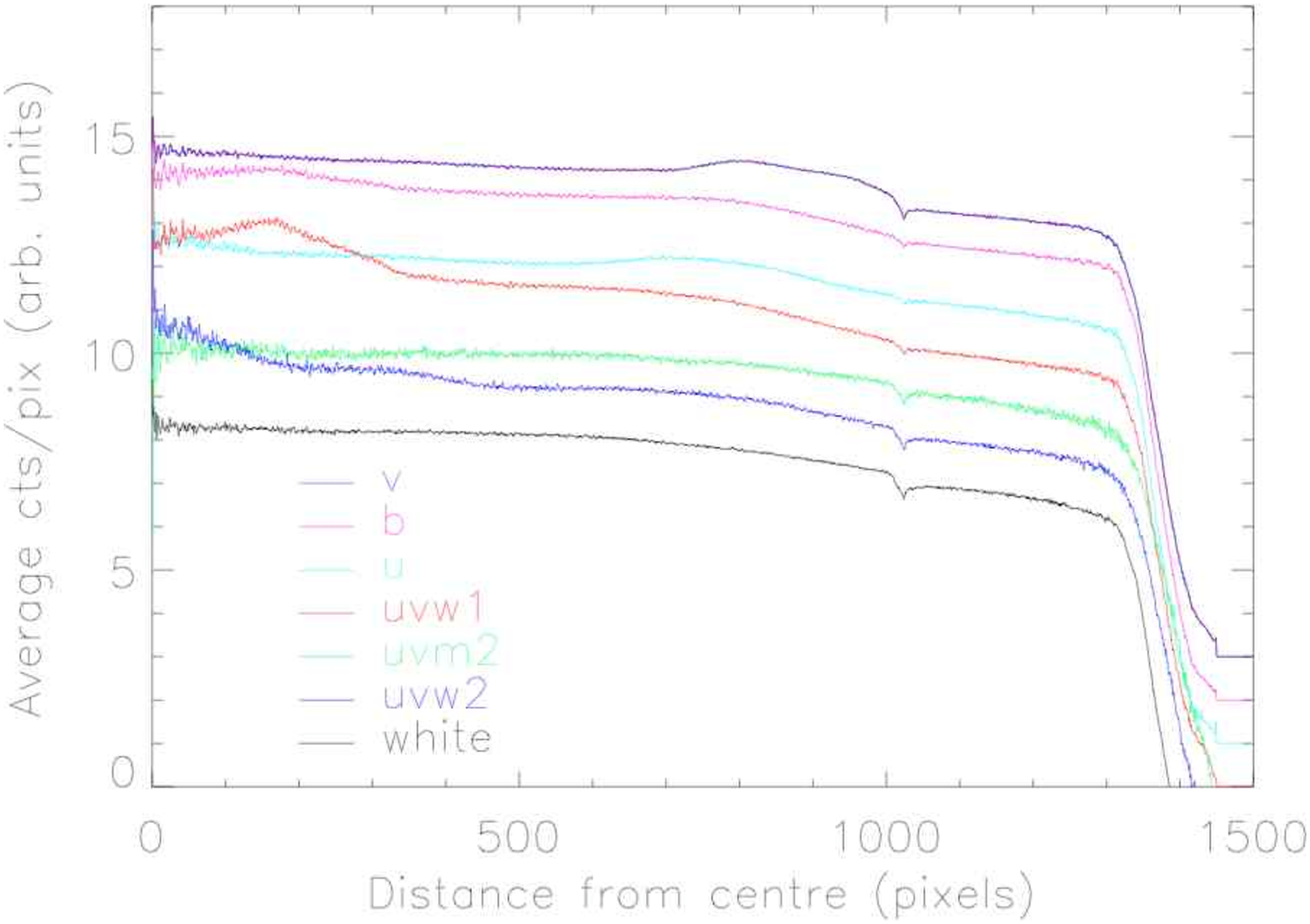}
\caption{Radial profiles of the scattered light. The curves are each offset by 1 unit on the y axis. For all filters the scattered light reduces towards the edges. The rings show up as small undulations on an otherwise smooth slope. The dips at a radius of 1024 pixels are due to the radius meeting the edge of the image where there are narrow strips of  about 20 pixels wide with markedly reduced counts. } 
\label{FigRadial}
\end{centering}
\end{figure}

Scattered light within the detector causes additional low level background, as well as some image artifacts. The scattered light has two causes: the first of these is due to starlight internally reflecting within the detector window causing two faint, out-of-focus ghost images of any bright star, one inside and one outside the primary image in the radial direction. The second effect is due to reflection of off-axis diffuse sky background light into the image from part of the detector housing which produces one or two very faint rings of enhanced background. Both these features are also seen in {\it XMM-OM Newton} \citep{MK2001} but the diffuse background scattered light is much reduced in the UVOT because of changes in the housing coating. The scattered light ring is at such a low level that it cannot be seen in individual images, and provided the background is measured close to the source, it should have no effect on photometry.

We have created images of the diffuse scattered light by taking full frame images from the UVOT archive and masking out all sources, ghost images, readout streaks etc., to leave just the background. Unbinned raw images for each filter were summed together using the background level and mask maps to normalise the values in each pixel. Fig.~\ref{FigScat} shows the diffuse scattered light for each filter. For the optical filters there are two rings: the outer one is centred on the middle of the image, but varies in radius from filter to filter, the inner one is offset from the centre at different positions for each filter.  The outer rings have between 1.2 and 2.2 per cent more counts per pixel compared with the region between the rings. The inner ring enhancement can be as much as 4 per cent. 

Fig.~\ref{FigRadial} shows the radial profiles of these scattered light images. The number of counts in each radial ring, 1 pixel wide, is divided by the number of pixels in that ring. The plots show that for all filters the scattered light reduces towards the edges. The rings show up as small undulations on an otherwise smooth slope. The dips at a radius of 1024 pixels are due to the radius meeting the edge of the image where there are narrow strips,  about 20 pixels wide, with reduced counts. These are caused by the onboard tracking, where UVOT autonomously determines the spacecraft drift using guide stars in the field of view and shifts the image to compensate. These narrow strips can be seen in the scattered light images (particularly the {\it white}) in Fig.~\ref{FigScat}.

The {\it white} and {\it uvm2} scattered light images do not show the distinct rings like the other filters.  It is not clear why this should be the case: possibly any features in the {\it white} filter are drowned by the high background count rate, while in the {\it uvm2} filter, the total background measured is lower than for any other filter. The {\it white} filter is rarely used in full frame mode because of telemetry restrictions, therefore there were only 36 {\it white} images contributing to this study. 

There are also visible some dark patches which are in identical positions for each filter. These patches do not appear on the flat fields made with the LED (see Section~\ref{SecSSS}), and therefore must be due to an optical element in the light path, such as the beam steering mirror. These could have a detrimental effect on photometry if a source were to land at these positions.  We are investigating this further and results will be documented on the {\it Swift} website\footnote{http://swift.gsfc.nasa.gov/}.

\subsection{Dark current}
The dark current has been monitored since pre-launch. It is very low (the most recent measurement being a count rate of $6.9\pm0.8\times10^{-5}$ s$^{-1}$~pix$^{-1}$) and has not changed throughout the mission. To enable direct comparison with the count rates given in Table~\ref{TabSky} this is equivalent to a count rate of 0.02 in a 5 arcsec aperture.  It can therefore be considered to be negligible compared with the other sources of background counts, except in the {\em uvm2} filter.

\section{Conclusions}
This paper substantially improves the calibration of the UVOT and compliments the photometric calibration already covered in Paper~I. This paper also discusses more fully some issues raised in that paper. 

We have measured the Point Spread Function out to a large radius (30 arcsec) for each filter to enable, for example, reliable surface photometry. We have also described how binning, rotation and orbital variations affect the PSF FWHM, and that photometry is not affected by the variations provided a big enough aperture is used.

Coincidence loss was described for a point source in Paper~I. Here we have extended the study to extended sources, and sources in regions of high background, presenting a model which not only shows us the limitations of the current coincidence correction, but also gives us the tools to improve it in the future. We also described the effect of mod-8 noise and how it can be removed without degrading photometric accuracy. 

We have measured the sensitivity variation over the detector and have constructed an effective correction for each filter that enables us to achieve a photometric response uniform to 1--2 per cent over the entire detector area. 

The positional accuracy of UVOT astrometry (after aspect correction) and distortion has been compared with the International Celestial Reference System and shown to be accurate to better than 0.42 arcsec (90 per cent confidence) including systematic and random errors.

Lastly, we have measured the range of observed background count rates in all filters and the dependencies on the Earth limb and Galactic latitude. We have discussed the sources of the background from both astrophysical and instrumental causes.

This extended calibration enables a more consistent use of UVOT data over the whole field of view. The work on the PSF, coincidence loss and astrometry are of particular use in the analysis of extended objects.  

\section*{Acknowledgments}

{\it Swift} UVOT was designed and built in collaboration between MSSL, PSU, SwRI, Swales Aerospace and GSFC, and was launched by NASA. We would like to thank all those involved in the continued operation of UVOT at PSU, MSSL and GSFC and those involved in the data processing and the writing of the analysis software.  This work is supported at MSSL by funding from STFC and at PSU by NASA's Office of Space Science through grant NAG5-8401 and NAS5-00136. We acknowledge the use of public data from the {\it Swift} archive.

\label{lastpage}


\begin{thebibliography}{99}
\bibitem[Barthelmy {\etal}(2005)]{BS2005}
	Barthelmy S.~D., Barbier L.~M., Cummings J.~R. et~al., 2005, Space Sci. Rev., 120, 143
\bibitem[Bertin \& Arnouts (1996)]{Bertin}Bertin E., Arnouts S., 1996, A\&AS 117,  393
\bibitem[Breeveld {\etal}(2005)]{spie1}
	Breeveld A.~A., Poole T.~S., James C.~J.  et~al., 2005, Proc. SPIE, 5898, 391
\bibitem[\protect\citeauthoryear{Brown, et al.}{2009}]{Brown09}
Brown P. ~J., Holland S.~T., Immler S. et al.,  2009, AJ, 137, 4517
\bibitem[Burrows {\etal}(2005)]{BDN2005}
	Burrows D.~N., Romano P., Falcone A. et~al., 2005, Space Sci. Rev., 120, 165
\bibitem[Feissel \& Mignard (1998)]{Feissel1998}Feissel M., Mignard F., 1998, A\&A, 331, L33
\bibitem[Fey {\etal}(2004)]{ICRS}Fey A.~L., Ma C., Arias E.~F. et~al., 2004, AJ, 127, 3587
\bibitem[\protect\citeauthoryear{Fordham, Moorhead \& Galbraith}{2000}]{FMG00}
Fordham  J.~L.~A., Moorhead  C.~F.,  Galbraith R.~F., 2000, MNRAS, 312, 83
\bibitem[Fricke {\etal}(1988)]{Fricke1988} Fricke W., Schwan H., Lederle T. et~al., 1988, Ver\"{o}f. Astr. Rechen-Institut, No. 32
\bibitem[Gehrels {\etal}(2004)]{GN2004}
	Gehrels N., Chincarini G., Giommi P. et~al., 2004, ApJ, 611, 1005
\bibitem[Hoversten {\etal}(2009)]{Hoversten}Hoversten E.~A. Gronwall C., Vanden Berk D.~E. et~al., 2009, ApJ, 705, 1462
\bibitem[Ivanushkina {\etal}(2005)]{spie2}Ivanushkina
  M., Breeveld A.~A., Poole T.~S. et~al., 2005, Proc. SPIE, 5898, 371
\bibitem[Ivezic \etal (2007)]{Ivezic}Ivezic Z., Allyn Smith J., Miknaitis G. et~al., 2007, AJ, 134, 973
\bibitem[\protect\citeauthoryear{James}{2007}]{James07}James C.,  September 2007, XMM document: Retrieval of Lost Spectral Information, Fixed Pattern Noise and Co-Incidence Loss Simulations
\bibitem[\protect\citeauthoryear{Kawakami, et al.}{1994}]{K94}
Kawakami H., Bone D., Fordham J., Michel R., 1994, Nucl. Instrum. Methods Phys. Res. A, 348, 707
\bibitem[Kuin \& Rosen (2008)]{Kuin} Kuin N.~P.~M, Rosen S.~R., 2008, MNRAS, 383, 383
\bibitem[Landsman (2009)]{Landsman}Landsman W., 2009, SWIFT-UVOT-CALDB-09-R03:  Large Scale Sensitivity
\bibitem[Mason {\etal}(2001)]{MK2001}
	Mason K.~O., Breeveld A.~A., Much R. et~al., 2001, A\&A, 365, L36
\bibitem[Michel, Fordham \& Kawakami(1997)]{Michel}Michel R., Fordham J., Kawakami H., 1997, MNRAS, 292, 611
\bibitem[Mignard \& Fr\symbol{27} schl\'{e} (2000)] {Mignard2000}Mignard F., Fr\symbol{27}schl\'{e} M., 2000, A\&A, 354, 732
\bibitem[Monet {\etal}(2003)]{USNO-B}Monet D.~G., Levine S.~E., Canzian B. et~al., 2003, AJ, 125, 984
\bibitem[Moretti {\etal}(2007)]{Moretti}
	Moretti A., Perri M., Capalbi M.  et~al., 2007, Proc. SPIE, 6688, 14
\bibitem[Poole {\etal}(2008)]{Poole}Poole T.~S., Breeveld A.~A., Page M.~J. et~al., 2008, MNRAS, 383, 627-645 (Paper~I)
\bibitem[Poole (2007a)]{Poolea}Poole T.~S., 2007a, SWIFT-UVOT-CALDB-10-R03: Zodiacal Light
\bibitem[Poole (2007b)]{Pooleb}Poole T.~S., 2007b, SWIFT-UVOT-CALDB-07-R03: Galactic Diffuse Light
\bibitem[Roming {\etal}(2005)]{Roming05}Roming P.~W.~A., Kennedy T.~E., Mason K.~O. et~al., 2005, Space Sci. Rev., 120, 95
\bibitem[Roming {\etal}(2009)]{Roming09}Roming P.~W.~A., Koch T.~S., Oates S.~R. et~al., 2009, ApJ, 690, 163
\bibitem[Stetson (1987)]{Stetson1987}Stetson P.~B., 1987, PASP, 99, 191
\bibitem[Stetson (2000)]{Stetson2000}Stetson P.~B., 2000, PASP, 112, 925
\bibitem[Stetson, McClure \& VandenBerg (2004)]{Stetson}Stetson P.~B., McClure R.~D., VandenBerg D.~A., 2004, PASP, 116, 1012
\bibitem[Zaritsky {\etal}(2002)]{Zaritsky}Zaritsky D., Harris J., Thompson I.~B., Grebel E.~K., Massey P., 2002, AJ, 123, 855.
\end{thebibliography}
\end{document}